\documentclass[journal]{IEEEtran}
\usepackage{amsmath,amsfonts}
\usepackage{amssymb}
\usepackage{amsthm}
\usepackage{makecell}
\usepackage{tabularx}
\usepackage{algorithmic}
\usepackage{algorithm}
\usepackage{array}
\usepackage[caption=false,font=normalsize,labelfont=sf,textfont=sf]{subfig}
\usepackage{graphicx}
\usepackage[justification=centering]{caption}
\usepackage{mathrsfs}
\usepackage{textcomp}
\usepackage{stfloats}
\usepackage{url}
\usepackage{bm}
\usepackage{cite}
\usepackage{color,xcolor}
\usepackage{verbatim}
\usepackage{graphicx}
\usepackage{pstricks}
\usepackage{epstopdf}
\usepackage{bm,bbm}
\newtheorem{theorem}{Theorem}
\newtheorem{lemma}{Lemma}
\newtheorem{proposition}{Proposition}
\newtheorem{corollary}{Corollary}

\newtheorem{remark}{Remark}

\def\BibTeX{{\rm B\kern-.05em{\sc i\kern-.025em b}\kern-.08em
		T\kern-.1667em\lower.7ex\hbox{E}\kern-.125emX}}
\usepackage{balance}

\begin{document}
	\title{Gaussian Arimoto--Blahut Algorithm for Capacity Region Calculation of Gaussian Vector Broadcast Channels}
	\author{Tian~Jiao, Yanlin~Geng,~\IEEEmembership{Member,~IEEE}, Anthony Man-Cho So,~\IEEEmembership{Fellow,~IEEE}, Yonghui~Chu, and Zai~Yang,~\IEEEmembership{Senior~Member,~IEEE}
		\thanks{A preliminary version of this work was presented in part at the IEEE International Symposium on Information Theory (ISIT), 2024 \cite{jiao2024blahut}. \emph{(Corresponding author: Zai~Yang.)}}
		\thanks{T.~Jiao, Y. Chu, and Z. Yang are with the School of Mathematics and Statistics, Xi'an Jiaotong University, Xi'an 710049, China (e-mail: tianjiao@stu.xjtu.edu.cn, chuyonghui@stu.xjtu.edu.cn,  yangzai@xjtu.edu.cn).} 
		\thanks{Y.~Geng is with the State Key Laboratory of ISN, Xidian University, China (e-mail: ylgeng@xidian.edu.cn).}
		\thanks{A. M.-C. So is with the Department of Systems Engineering and Engineering Management, The Chinese University of Hong Kong, Hong Kong SAR, China (e-mail: manchoso@se.cuhk.edu.hk).}
	}
	
	\maketitle
	
	\begin{abstract}
		This paper is concerned with the computation of the capacity region of a continuous, Gaussian vector broadcast channel (BC) with covariance matrix constraints. 
		Since the decision variables of the corresponding optimization problem are Gaussian distributed, they can be characterized by a finite number of parameters. Consequently, we develop new Arimoto--Blahut (AB)-type algorithms that can compute the capacity without discretizing the channel.			 
		First, by exploiting projection and an approximation of the Lagrange multiplier, which are introduced to handle certain positive semidefinite constraints in the optimization formulation, we develop the Gaussian AB algorithm with projection (GAB-P).
		Then, we demonstrate that one of the subproblems arising from the alternating updates admits a closed-form solution. Based on this result, we propose the Gaussian AB algorithm with alternating updates (GAB-A) and establish its convergence guarantee. 
		Furthermore, we extend the GAB-P algorithm to compute the capacity region of the Gaussian vector BC with both private and common messages.
		All the proposed algorithms are parameter-free. 
		Lastly, we present numerical results to demonstrate the effectiveness of the proposed algorithms.		
	\end{abstract}
	

	\section{Introduction}
\IEEEPARstart{M}{ulti-antenna} downlink systems, particularly massive multiple-input multiple-output (MIMO), are essential to 5G/B5G/6G networks \cite{jindal2008antenna, shirani2010mimo, kobayashi2011training, ghazanfari2021model}. Within the framework of network information theory, the fundamental performance limits of such systems are captured by the capacity region of the vector broadcast channel (BC) model.

	As a fundamental and commonly used class of BCs, Gaussian vector BC has attracted wide attention \cite{caire2003achievable, vishwanath2003duality, viswanath2003sum, yu2004sum}. 
	The authors of \cite{caire2003achievable} derived the sum capacity of the Gaussian vector BC with two receivers, each equipped with a single antenna, by exploiting dirty paper coding \cite{costa1983writing} and Sato’s outer bound \cite{sato1978outer}. 
	The sum capacity of the BC is independently obtained in  \cite{vishwanath2003duality} and \cite{viswanath2003sum}  by utilizing the duality between the capacity region of the multiple-access channel (MAC) and the dirty paper coding region of the BC. The conclusions in \cite{caire2003achievable} are generalized to the sum capacity of a vector BC with an arbitrary number of transmit antennas and users in \cite{yu2004sum}, where each user is equipped with multiple receive antennas.
	
	To address optimization problems related to the Gaussian vector BC, an effective approach is to exploit the BC-MAC duality \cite{rashid1998transmit, jindal2005sum, yu2006sum, yu2006uplink, yu2007transmitter, zhang2012gaussian}.
	The total power minimization problem for BC with received signal-to-interference-plus-noise-ratio (SINR) constraints is solved in \cite{rashid1998transmit} by converting the non-convex BC problem into a convex MAC problem using the BC-MAC duality. It is shown in \cite{jindal2005sum} that the sum capacity of the BC is equivalent to that of the dual MAC under a single transmit power constraint. The authors of \cite{yu2006sum} proposed to compute the sum capacity of the Gaussian vector BC via a Lagrangian dual decomposition technique. The authors of \cite{yu2006uplink, yu2007transmitter} showed that arbitrary boundary points of the BC capacity region can be obtained by solving a dual minimax optimization problem in the MAC setting, either under a sum power constraint or a set of linear power constraints. The weighted sum rate of the Gaussian vector BC under multiple linear transmit covariance constraints is further characterized in \cite{zhang2012gaussian} based on the BC-MAC duality. However, the aforementioned papers did not completely resolve the capacity region problem of the Gaussian vector BC.
	
	The capacity region of the vector BC has been characterized in \cite{weingarten2006capacity, geng2014capacity}. Specifically, the authors of \cite{weingarten2006capacity} established the capacity region of the two-receiver Gaussian vector BC with private messages, demonstrating that a pair of inner and outer bounds yields identical regions. However, this argument could not be generalized to the cases of Gaussian vector BC with both private and common messages. 
	The authors of \cite{geng2014capacity} developed a method to establish the optimality of Gaussian auxiliary random variables in multiterminal information theory problems and applied it to show that Marton's inner bound achieves the capacity region of the two-receiver Gaussian vector BC with both private and common messages.
	
	With the characterization of the capacity region of the Gaussian vector BC, the calculation of the capacity region with autocorrelation matrix constraints has garnered renewed interest in recent years \cite{lau2022uniqueness,yao2023globally}.
	The authors of \cite{lau2022uniqueness} showed that the optimization problem corresponding to the capacity region of the Gaussian vector BC has a unique local (hence global) maximizer and provided a path to the optimal point. However, it is unclear how to utilize the path for algorithm design since its expression depends on the optimal point. Within the framework of the difference-of-convex algorithm (DCA), the authors of \cite{yao2023globally} proposed the DCProx algorithm to solve the optimization problem by iteratively solving a series of convex subproblems using the primal-dual proximal algorithm with Bregman distance. They also proved that the proposed algorithm converges to the optimal point at a linear rate.   
	It should be noted that the capacity region of the Gaussian vector BC with covariance matrix constraints may not be dual to that of MAC and thus may not be amenable to methods that are developed for computing the capacity region of MAC.
	
	The Arimoto--Blahut (AB) algorithm, developed independently by Arimoto \cite{arimoto1972algorithm}  and Blahut  \cite{blahut1972computation}, is a widely used method for calculating channel capacity in information theory. Specifically, to calculate the channel capacity of a point-to-point channel $p(y|x)$, i.e.,
	$$\max_q \left\{I(X;Y) =\int q(x) p(y|x) \ln \frac{ q(x|y) }{ q(x) } {\rm d}x {\rm d}y\right\},$$
	the AB algorithm replaces the conditional probability mass function $q(\cdot | \cdot)$ by a free variable $Q(\cdot | \cdot)$ and then maximizes the objective function over $q(\cdot)$ and $Q(\cdot | \cdot)$ alternatingly. The authors of \cite{liu2022blahut} developed AB-type algorithms to evaluate the supporting hyperplanes of the superposition coding region and those of the UV outer bound, as well as the sum-rate of Marton’s inner bound. However, the classical AB algorithm is only applicable to discrete channels and does not apply to continuous channels. A common alternative approach is to discretize the continuous channel first and then apply the AB algorithm to calculate its capacity approximately. Besides the discretization error, the computational complexity of the algorithm increases dramatically with the fineness of the discretization.
	
	In this paper, we focus on calculating the capacity region of the Gaussian vector BC. 
	For the capacity region of the Gaussian vector BC with private messages, we derive an equivalent formulation of the corresponding optimization problem to simplify the set of constraints. 
	Within the framework of the AB algorithm, we transform the distribution optimization problem into an optimization problem concerning the covariance matrix by leveraging the property of Gaussian distribution.
	We apply projection and an approximation of the Lagrange multiplier, which are introduced to handle certain positive semidefinite constraints in the formulation, to develop the Gaussian AB algorithm with projection (GAB-P).
	We then examine one of the subproblems arising from the alternating updates. By exploiting the structure of its stationary point set, we derive the Gaussian AB algorithm with alternating updates (GAB-A) and establish its convergence guarantee.
	For the capacity region of the Gaussian vector BC with both private and common messages, we adopt the
	GAB-P algorithm to solve the corresponding optimization subproblems, thereby giving rise to the extended GAB-P algorithm (EGAB-P).
	
	The rest of the paper is organized as follows. Section \ref{secmr} develops the GAB-P and GAB-A  algorithms for calculating the capacity region of the Gaussian vector BC with private messages. Section \ref{secom} generalizes the proposed algorithms to the case of the Gaussian vector BC with both private and common messages. Section \ref{secop} evaluates the performance of the proposed algorithms through numerical simulations. Section \ref{secco} concludes the paper.
	
	\paragraph*{Notation}
	We denote  by $\mathbb{S}$ the set of symmetric matrices, by $\mathbb{S}_+$ the set of symmetric positive semidefinite (PSD) matrices, by $\mathbb{S}_{++}$ the set of symmetric positive definite (PD) matrices, and by $\mathbb{S}_K$  the set $\{M\in \mathbb{S}: M\preceq K\}$ for a given $K \in \mathbb{S}$, where $M\preceq K$ means that $K - M \in \mathbb{S}_+$. Given $K \in \mathbb{S}_+$, we denote by $|K|$ the determinant of $K$ and write $K \succ 0$ to mean that $K \in \mathbb{S}_{++}$. We denote by $\mathbb{A}^c$ the complement of the set $\mathbb{A}$.
	Let $X$ be a continuous random vector. The differential entropy of $X$ is denoted as $h(X)$. We write $ X\sim \mathcal{N}\big(\mu,\Sigma\big)$ to mean that the random vector $X$ is normally distributed with mean $\mu$ and  covariance matrix $\Sigma$ and $ a \propto b$ to mean that $a$ is proportional to $b$. We denote by $g(x; \mu,\Sigma) \propto \exp\left( -\frac{1}{2} (x-\mu)^T \Sigma^{-1} (x-\mu) \right)$ the probability density function (pdf) of $\mathcal{N}(\mu, \Sigma)$. We denote by $\text{diag}(a_1,a_2,\ldots, a_n)$ the diagonal matrix with diagonal elements $(a_1,a_2,\ldots , a_n)$, by $I$ the identity matrix, and by $\|\cdot\|_2$ the $\ell_2$-norm. We denote by $D(\cdot \,\|\, \cdot)$ the Kullback--Leibler divergence.

	\section{Gaussian Arimoto--Blahut Algorithms for Gaussian Vector BC with Private Messages}\label{secmr}
	In this section, we focus on developing algorithms to compute the capacity region of Gaussian vector BC with private messages. 
	Based on the AB algorithm framework,  we transform the corresponding infinite-dimensional problem into an equivalent finite-dimensional one by exploiting the properties of Gaussian distribution and propose two Gaussian AB algorithms.
		\subsection{Classical AB Algorithm}\label{CAB}
		Before introducing the Gaussian AB algorithms, we review the classical AB algorithm, which was originally developed to compute the channel capacity of discrete memoryless point-to-point channels. Recall that for a discrete memoryless point-to-point channel with input $X$ and output $Y$, its channel capacity is given by 
		\begin{align} \label{oppo1}
			C = \max_{q(x)} I(X;Y), 
		\end{align}
		where $q(x)$ is the input distribution.
		Based on the mutual information formula
		\begin{align*}
			I(X;Y)=\sum_{x,y}  q(x) p(y|x) \ln\frac{q(x|y)}{q(x)}:=F(q),
		\end{align*}
		consider the expression
		\begin{align*}
			&\sum_{x,y}  q(x) p(y|x) \ln\frac{Q(x|y)}{q(x)} {\rm d}x{\rm d}y
			\\
			=&\sum_{x}  q(x) \big(\sum_{y} p(y|x) \ln Q(x|y)-\ln q(x)\big):= F(q,Q),
		\end{align*}
		where $q(x|y)$ is replaced by a new variable $Q$. The following result serves as a starting point of the classical AB algorithm.
		\begin{lemma}(\cite[Theorem 1]{blahut1972computation})\label{th1AB}
			The following properties hold.
			
			(a) $C = \max_{q(x)}\max_{Q(x|y)} F(q,Q)$.
			
			(b) For fixed $q(x)$, the function $Q \mapsto F(q,Q)$ is maximized by $Q(x|y)=q(x|y)$.
			
			(c) For fixed $Q(x|y)$, the function $q \mapsto F(q,Q)$ is maximized by 
			\begin{align}\label{ABq}
				q(x) = \frac{e^{\sum_{y} p(y|x) \ln Q(x|y)}}{\sum_{x} e^{\sum_{y} p(y|x) \ln Q(x|y)}}.
			\end{align} 		
		\end{lemma} 
		
		 Based on the above result, we see that while
		the joint maximization of $F$ over $(q,Q)$ is intractable, the maximization of $F$ over one variable (either $q$ or $Q$) admits a closed-form solution once the other variable is fixed. This leads to the development of the AB algorithm, which performs an alternating update of the two subproblems. However, it is noted that this algorithm is specifically designed for discrete distributions. For continuous distributions, it is usually necessary to convert them into discrete distributions through quantization before applying the AB algorithm. Unfortunately, the problem size grows exponentially with the discretization degree.
		To circumvent this difficulty, we observe that for pdfs that can be characterized by a finite number of parameters, we can update those parameters via the AB algorithm, thereby avoiding discretization when calculating the channel capacity. In the following, we develop Gaussian AB algorithms based on this idea for capacity region computation of Gaussian vector BCs.
	
	\subsection{Problem Formulation}\label{ASEC}

		Consider the Gaussian vector BC with covariance matrix constraints 
		\begin{align*} 
			Y_1 &= X + Z_1, \\ 
			Y_2 &= X + Z_2, 
		\end{align*}
		where $0\preceq \mathbb {E}[{XX}^T]\preceq  K$, ${Z_1}\sim \mathcal{N}(0,\Sigma_1)$, ${Z_2}\sim  \mathcal{N}(0,\Sigma_2)$, and $K \succeq 0$, $\Sigma_1\succ 0$, and $\Sigma_2\succ 0$ are fixed.
		It is shown in \cite{geng2014capacity} that the supporting hyperplane of the capacity region $\mathcal{C}$ for the Gaussian vector BC with private messages $p(y_1, y_2|x)$ and parameter $\lambda > 1$ can be characterized by the formula
		\begin{align}\label{eq1}
			\max_{(R_1,R_2) \in \mathcal{C} \atop {X}:\mathbb {E}[{XX}^T]= K  }R_1+ \lambda  R_2  &=\max_{q(v,x) } I({X};{Y_1}|{V})+\lambda I\left({V} ; {Y_2} \right)\notag\\
			&=I({X}_*;{Y_1}|{V}_*)+\lambda I\left({V}_* ; {Y_2} \right),
		\end{align}
		where $R_1,R_2$ are the message rates, $q(v,x)$ represents the joint pdf of $({V}, X)$, and ${V}_*, {U}_*, X_* $ denote the optimal $V, U, X$ with ${V}_*\sim   \mathcal{N}(0, K_V )$, ${U}_*\sim  \mathcal{N}(0, K_U )$ for some $K_U, K_V \succeq 0$ being independent and satisfying $X_* = V_*+U_*$ and $K= K_V+K_U$. 
		Since the capacity region $\mathcal{C}$ is a closed and bounded convex set, it can be fully characterized by the intersection of its supporting hyperplanes, i.e., the intersection of all supporting hyperplanes $R_1+ \lambda  R_2$ with $\lambda>0$. For $\lambda >1$, the supporting hyperplane can be obtained by solving the optimization problem \eqref{eq1}. For $\lambda <1$, the supporting hyperplane can be obtained similarly by exploiting the symmetry of $Y_1$ and $Y_2$ in the outer bound. 
		For $\lambda =1$, the supporting hyperplane can be obtained by utilizing the continuity of $\lambda \mapsto \max_{(R_1,R_2) \in \mathcal{C}} R_1 + \lambda R_2$ at $\lambda = 1$ \cite[Remark 9]{geng2014capacity}.

		Building upon the fundamental assumption of discrete memoryless broadcast channels, where the signal source $(U, V)$ is independent of channel noise $(Z_1, Z_2)$, we apply the differential entropy formula for Gaussian distributions \cite{el2011network} to the mutual information expression in \eqref{eq1} to obtain the following formulas:
		\begin{align}
			I\left({V}_* ; {Y_2} \right) &= h(Y_2)-h(Y_2|V_*)\notag\\&=\frac{1}{2}\big(\ln|K+\Sigma_2|-\ln|K_U+\Sigma_2|\big),\label{f1} 
		\end{align}
		\begin{align}
			I({X}_*;{Y_1}|{V}_*) &= h(Y_1|V_*)-h(Y_1|X)\notag\\&=\frac{1}{2}\big(\ln|K_U+\Sigma_1|-\ln|\Sigma_1|\big).\label{s1}
		\end{align} 
		Inserting \eqref{f1}--\eqref{s1} into the objective function of the optimization problem \eqref{eq1}, we obtain 
		\begin{align*}
			&I({X}_*;{Y_1}|{V}_*)+\lambda I\left({V}_* ; {Y_2} \right) = \frac{1}{2}\big( \ln|K_U+\Sigma_1|-\ln|\Sigma_1|\\&\ +\lambda\ln|K+\Sigma_2| -\lambda\ln|K_U+\Sigma_2|\big).
		\end{align*}		
		Since the quantity $\lambda\ln|K+\Sigma_2|-\ln|\Sigma_1|$ is a constant, the optimization problem (\ref{eq1}) is equivalent to 
		\begin{align} \label{o1}
			\max_{ 0 \preceq K_U \preceq K }\ln|K_U+\Sigma_1|-\lambda\ln|K_U+\Sigma_2|,
		\end{align} 
		where $K \succeq 0$, $\Sigma_1\succ 0$,  $\Sigma_2\succ 0$, and $\lambda >1$ are fixed. The optimization problem \eqref{o1}  is non-convex since the objective function is the difference of two concave functions.

	In the following, we show that the matrix $K$ in the optimization problem \eqref{o1} can be replaced by the identity matrix without loss of generality. To begin, let $K = PLP^T$ be the eigen-decomposition of $K$, where $P$ is an orthogonal matrix and $L=\text{diag}(l_1,\ldots,l_n)$ is a diagonal matrix with 
	$l_1\geq l_2\geq \cdots\geq l_r> l_{r+1}= \cdots= l_n=0$ and $r={\rm rank}(K)\leq n$. Let $\tilde{L} = \text{diag}(1/{\sqrt{l_1}}, \ldots, 1/{\sqrt{l_r}}, 1,\ldots,1)$ and  
	define
	\begin{equation} \label{kuk}
		\tilde{K} = \tilde{L}P^T, \quad \tilde{K}_U = \tilde{K} K_U \tilde{K}^T, \quad \tilde{\Sigma}_j = \tilde{K} \Sigma_j \tilde{K}^T, \,\,\, j=1,2,
	\end{equation}
	and $ I_r = \tilde{K} K \tilde{K}^T = {\rm diag}( \underbrace{1,\ldots,1}_{r},0,\ldots,0 ).$
	Based on the block structure of $I_r$, we consider the following block decomposition of $\tilde{K}_U$, $\tilde{\Sigma}_1$, and $\tilde{\Sigma}_2$:
	\begin{align}\label{decom}
		\tilde{K}_U  = \begin{bmatrix}
			A_U & B_U \\ B_U^T &C_U 
		\end{bmatrix},	\quad		
		\tilde{\Sigma}_{1} = \begin{bmatrix}
			A_1 & B_1 \\ B_1^T &C_1 
		\end{bmatrix},\quad
		\tilde{\Sigma}_{2} = \begin{bmatrix}
			A_2 & B_2 \\ B_2^T &C_2 
		\end{bmatrix}.
	\end{align}
	We can then prove the following proposition:
	\begin{proposition}\label{prop1}
		The optimization problem \eqref{o1} is equivalent to 
		\begin{align}\label{eqform}
			\max_{ 0 \preceq A_U \preceq I }\ln|A_U+\hat{{\Sigma}}_{1}|-\lambda\ln|A_U+\hat{{\Sigma}}_{2}|,
		\end{align}
		where 
		\begin{align}\label{kj}
			\hat{{\Sigma}}_{1} = A_1 - B_1C_1^{-1} B_1^T,
			\quad\hat{{\Sigma}}_{2} = A_2 - B_2C_2^{-1} B_2^T.
		\end{align} 
	\end{proposition}
	\begin{IEEEproof}	
		Since $\tilde{K}$ is invertible, we have
		\begin{align}
			\ln |K_U+\Sigma_j|
			=& \ln |\tilde{K}^{-1}\tilde{K}(K_U+\Sigma_j)\tilde{K}^T(\tilde{K}^T)^{-1}| \notag\\=&\ln |\tilde{K}_U +  \tilde{\Sigma}_{j}|+ c\label{eq3}
		\end{align}
		for some constant $c$ that is independent of the variables.\footnote{We denote by $c$ a generic constant whose value may change from appearance to appearance.}
		Inserting (\ref{eq3}) into (\ref{o1}), we see that the latter is equivalent to
		\begin{align*}
			\max_{ 0 \preceq \tilde{K}_U \preceq I_r }\ln|\tilde{K}_U +  \tilde{\Sigma}_{1}|-\lambda\ln|\tilde{K}_U +  \tilde{\Sigma}_{2}|.
		\end{align*}
		According to the column inclusion property  of PSD matrices \cite{horn2012matrix} and the fact that the principal minors of PSD matrices are also PSD, the constraint $0\preceq \tilde{K}_U\preceq I_r$ is equivalent to $0\preceq A_U \preceq I$, $B_U=0$, and  $C_U=0$. Furthermore, we have
		\begin{align*}
			&\ln |\tilde{K}_U +  \tilde{\Sigma}_{j}| 
			= \ln \left|\begin{bmatrix}
				A_U+A_j & B_j \\ B_j^T &C_j
			\end{bmatrix} \right|\\=& \ln|C_j| + \ln | A_U+A_j - B_jC_j^{-1} B_j^T|,
		\end{align*}
		and the Schur complement theorem \cite{zhang2006schur} guarantees that  $A_j - B_jC_j^{-1} B_j^T\succ 0$ when $\tilde{\Sigma}_{j}\succ 0$. This completes the proof.
	\end{IEEEproof}
	
	To map $A_U$ back to $K_U$, we set 
	\begin{align}\label{kn}
		K_U=\tilde{K}^{-1}\begin{bmatrix}
			A_U & 0 \\ 0 &0
		\end{bmatrix} (\tilde{K}^T)^{-1}=\tilde{K}^{\dag}A_U{(\tilde{K}^{\dag})}^T,
	\end{align}
	where $\tilde{K}^{\dag}=P_{:,1:r} \text{diag}({\sqrt{l_1}}, \ldots, {\sqrt{l_r}})$ and $P_{:,1:r}$ is the $n\times r$ matrix composed of the first $r$ columns of $P$.
	
	Based on the preceding analysis, we proceed to derive an AB-type algorithm for solving the optimization problem \eqref{eq1} with $K=I$. According to the capacity region characterization in (\ref{eq1}), we formulate the mutual information expression as (we drop the subscript $*$ to simplify notation)
	\begin{align}\label{fq}
		F(q)
		= &\lambda I(V;Y_2) + I(X;Y_1|V) \notag\\
		\stackrel{(a)}{=} &\lambda (h(V)-h(V|Y_2)) + I(U;Y_1|V) \notag\\ 
		\stackrel{(b)}{=} &\lambda (h(V)-h(V|Y_2)) + h(U)-h(U|V,Y_1)\notag\\
		= &\int  q(u)q(v) p(y_1,y_2|u+v) \Big( \lambda \ln\frac{q(v|y_2)}{q(v)}-\ln q(u)\notag\\ &+\ln q(u|v,y_1)\Big) {\rm d}u{\rm d}v{\rm d}y_1{\rm d}y_2,
	\end{align}
	 where $q(u)$ and $q(v)$ are the pdfs of the Gaussian vectors $U$ and $V$, respectively, $(a)$ holds by $X=V+U$, and $(b)$ holds by the independence of $U$ and $V$.
	Upon replacing the conditional probabilities $q(v|y_2)$ and $q(u|v,y_1)$ by the free variables $Q(v|y_2)$ and $Q(u|v,y_1)$, respectively, we define, with a slight abuse of notation, the quantity
	\begin{align}\label{fqQ}
		&F(q,Q)\notag\\
		= &\int  q(u)q(v) p(y,y_2|u+v) \Big( \lambda \ln\frac{Q(v|y_2)}{q(v)}-\ln q(u)+\ln Q(\notag\\ &\quad u|v,y_1)\Big) {\rm d}u{\rm d}v{\rm d}y_1{\rm d}y_2\notag\\
		= &\int  q(u)q(v) p(y_1|u+v) \big( \ln Q(u|v,y_1) -\ln q(u)\big) {\rm d}u{\rm d}v{\rm d}y_1\notag\\ & 
		 + \lambda\int q(u)q(v) p(y_2|u+v) \big( \ln Q(v|y_2) - \ln q(v) \big) {\rm d}u{\rm d}v{\rm d}y_2\notag \\
		= &\int q(u)\big(d_U\big[Q\big](u)-\ln q(u)\big) {\rm d}u + \lambda \int q(v) \big(d_V\big[Q\big](v)-\notag\\&\quad \ln q(v)\big) {\rm d}v,
	\end{align}
	where 
	\begin{align*}
		d_U\big[Q\big](u) 
		&=  \int q(v) p(y_1|u+v) \ln Q(u|v,y_1) {\rm d}v{\rm d}y_1 ,\\
		d_V\big[Q\big](v) 
		&= \int q(u) p(y_2|u+v) \ln Q(v|y_2) {\rm d}u{\rm d}y_2.
	\end{align*}	
	Consider first the problem of maximizing $F(q,\cdot)$ over all pdfs $Q$. We have the following theorem:
	\begin{theorem}\label{th1}
		Given the pdfs $q(u)$ and $q(v)$, the maximizing pdf $Q[q]$ of $F(q,\cdot)$ satisfies
		\begin{align}\label{Qq}
			Q[q](v|y_2) =q(v|y_2), \quad Q[q](u|v,\,y_1) =q(u|v,\,y_1).
		\end{align}
		Further, since  $V\sim  \mathcal{N}(0,K_V), U\sim  \mathcal{N}(0,K_U)$, and $K_V+K_U=I$, we get
		\begin{align}
			Q[q](v|y_2)&=   g(v; Ay_2, W_1),\label{Qvz}\\
			Q[q](u|v,\,y_1)&= g(u; B(y_1-v), W_2),\label{Quyv}
		\end{align}
		where $A=K_V(I+\Sigma_2)^{-1}$, $W_1=K_V-K_V(I+\Sigma_2)^{-1}K_V$, $B=K_U(K_U+\Sigma_1)^{-1}$, and $W_2=K_U-K_U(K_U+\Sigma_1)^{-1}K_U$. 		
	\end{theorem} 
	\begin{IEEEproof}
		Based on \eqref{Qq} and the definition of Kullback--Leibler divergence, we have 
		\begin{align*}
			&F(q,Q[q])-F(q,Q)=F(q)-F(q,Q)\\
			=&\int  q(v,y_1)D\big(q(u|v,y_1)\parallel Q(u|v,y_1)\big){\rm d}v{\rm d}y_1+\lambda\int  q(y_2)\notag\\ &\quad D\big(q(v|y_2)\parallel Q(v|y_2)\big){\rm d}y_2  \ge 0.
		\end{align*}
		Besides, the expressions of $ Q[q](u|v,y_1)$ and $Q[q](v|y_2)$ can be obtained from the conditional distribution of a multivariate Gaussian distribution since $V\sim  \mathcal{N}(0,K_V), U\sim  \mathcal{N}(0,K_U)$, and $K_V+K_U=I$. Specifically, we have $Y_1\sim  \mathcal{N}(0,I+\Sigma_1)$ and $Y_2\sim  \mathcal{N}(0,I+\Sigma_2)$. According to the formula of conditional distribution of a Gaussian distribution \cite{eaton1983multivariate}, the distribution of $V$ conditional on $Y_2=y_2$ is Gaussian, i.e., 
		$$V | (Y_2=y_2)\sim  \mathcal{N}(K_V(I+\Sigma_2)^{-1}y_2,K_V-K_V(I+\Sigma_2)^{-1}K_V).$$
		Similarly, the distribution of $U$ conditional on $V=v$ and $Y_1=y_1$ is Gaussian, i.e., 
		\begin{align*}
			U | (V=v, \, Y_1=y_1)\sim  \mathcal{N}&\big(K_U(K_U+\Sigma_1)^{-1}(y_1-v),K_U-\notag\\ &\quad K_U(K_U+\Sigma_1)^{-1}K_U\big).
		\end{align*}
		This completes the proof.
	\end{IEEEproof}
	Making use of Theorem \ref{th1}, we can derive the following explicit expression of $F(q,Q[q])$:
	\begin{theorem}\label{th3}
		Under the conditions of Theorem \ref{th1} and assuming $K_U,K_V \succ 0$, we have
		\begin{align} \label{opquv}
			F\left(q,Q[q]\right) =& \int q(u) \left( -\frac{1}{2} u^T D_U u - \ln q(u) \right) {\rm d}u  \notag\\
			&+ \lambda \int q(v)\left( -\frac{1}{2} v^T D_V v - \ln q(v) \right) {\rm d}v + c,
		\end{align}
		where $D_U = (I-B)^TW_2^{-1}(I-B)$, $D_V = W_1^{-1}-A^T W_1^{-1}-W_1^{-1} A$ and $c$ is a constant.
	\end{theorem}
	\begin{IEEEproof}	
			Since $K_U,K_V \succ 0$, both 
			\begin{align*}
				&W_1=K_V-K_V(I+\Sigma_2)^{-1}K_V=(I+\Sigma_2-K_V)(I\\
				&\ +\Sigma_2)^{-1}K_V=(K_U+\Sigma_2)(I+\Sigma_2)^{-1}K_V
			\end{align*} and 
			\begin{align*}
				W_2=K_U-K_U(K_U+\Sigma_1)^{-1}K_U=\Sigma_1(K_U+\Sigma_1)^{-1}K_U
			\end{align*}
			are invertible, and thus $D_U$ and $D_V$ are well defined. 
		By combining the fact that $p(y_1|u+v)=g(y_1; u+v, \Sigma_1)$ with (\ref{Quyv}), we obtain
		\begin{align*}
			&2  \int  p(y_1|u+v) \ln Q[q](u|v,y_1) {\rm d}y_1 \\
			= &\  \mathbb E_{Y_1|u+v} \big[ - \big(u-B(Y_1-v))^T W_2^{-1} (u-B(Y_1-v)\big) \big] + c_1\\
			= & - (u-Bu)^T W_2^{-1} (u-Bu) - {\rm{tr}}(B^TW_2^{-1}B\Sigma_1)+ c_1,
		\end{align*}
		where  $c_1=-\ln(2\pi)^n|W_2|$ is a constant with $n$ being the dimension of $U$.
		Then, taking expectation over $V$, we get
		\begin{align*}
			&d_U\big[Q[q]\big](u) \\
			=&  - \frac12   u^T(W_2^{-1} - W_2^{-1} B - B^TW_2^{-1} + B^T W_2^{-1}B)u  + c_2\\
			=& - \frac12 u^T(I-B)^TW_2^{-1}(I-B)u+ c_2\\
			=& -\frac{1}{2} u^T D_U u + c_2,
		\end{align*}	
		where $c_2=\big(c_1 -{\rm{tr}}(B^TW_2^{-1}B\Sigma_1)\big)/2$ is a constant.	
		Similarly, by combining the fact that $p(y_2|u+v)=g(y_2; u+v, \Sigma_2)$ with \eqref{Qvz}, we obtain
		\begin{align*}
			&2  \int  p(y_2|u+v) \ln Q[q](v|y_2) {\rm d}y_2 \\
			= & \mathbb {E}_{Y_2|u+v} [ - (v-Ay_2)^T W_1^{-1} (v-Ay_2) ]+ c_3  \\=& -[v^T W_1^{-1}v-(u+v)^TA^TW_1^{-1}v-v^T W_1^{-1}A(u+v) ]\\ & -{\rm{tr}}\big(A^TW_1^{-1}A(u+v)(u+v)^T\big)
			-{\rm{tr}}(A^TW_1^{-1}A\Sigma_2)+ c_3,
		\end{align*}
		where $c_3=-\ln(2\pi)^n|W_1|$ is a constant with $n$ being the dimension of $V$. Then, taking expectation over $U$, we get 
		\begin{align*}
			&d_V\big[Q[q]\big](v)\\
			=& - \frac12 \big( v^T(W_1^{-1}-A^T W_1^{-1}-W_1^{-1} A+A^TW_1^{-1}A)v\\ &\ +{\rm{tr}}(A^TW_1^{-1}AK_U)\big)+ c_4,
		\end{align*}
		where $c_4=\big(c_3 -{\rm{tr}}(A^TW_1^{-1}A\Sigma_2)\big)/2$ is a constant.			
		From the above identity, we see that the variable $K_U$ appears in $d_V\big[Q[q]\big](v)$. 
		Now, define
		\begin{align*}
			\tilde{d}_V\big[Q[q]\big](v) =- \frac12 v^T(W_1^{-1}-A^T W_1^{-1}-W_1^{-1} A)v+ c_4.
		\end{align*}		
		Thus, we have 
		\begin{align*}
			d_V\big[Q[q]\big](v)=& \ \tilde{d}_V\big[Q[q]\big](v)- \frac12 v^TA^TW_1^{-1}Av \\ & -\frac12{\rm{tr}}(A^TW_1^{-1}AK_U)
		\end{align*}
		and 
		\begin{align*}
			&   \int q(v) d_V\big[Q[q]\big](v) {\rm d}v \notag\\
			=\ &    \mathbb {E}_{V} [ \tilde{d}_V\big[Q[q]\big](V)] - \frac12\mathbb {E}_{V} [ V^TA^TW_1^{-1}AV]\\ &- \frac12{\rm{tr}}(A^TW_1^{-1}AK_U)\notag\\
			\stackrel{(a)}{=}\ &     \mathbb {E}_{V} [ \tilde{d}_V\big[Q[q]\big](V)]- \frac12{\rm{tr}}(A^TW_1^{-1}A)\notag\\
			=\ &   \int q(v) \left(-\frac12v^T(W_1^{-1}-A^T W_1^{-1}-W_1^{-1} A)v+ c_5\right) {\rm d}v,\notag\\
			=\ & \int q(v) \left( -\frac{1}{2} v^T D_V v +c_5\right) {\rm d}v,
		\end{align*}
		where $(a)$ holds by $V\sim  \mathcal{N}(0,K_V), U\sim  \mathcal{N}(0,K_U), K_V+K_U=I$ and $c_5=c_4- \frac12{\rm{tr}}(A^TW_1^{-1}A)$ is a constant. Therefore, we get \eqref{opquv} with $c=c_2+\lambda c_5$, as desired.
	\end{IEEEproof}

		Similar to the derivation of Theorem \ref{th3}, below we give an alternative expression for $F\left(q,Q[q]\right)$.
		\begin{corollary}\label{copo1}
			Under the conditions of Theorem \ref{th1} and assuming $K_U,K_V \succ 0$, we have
			\begin{align} \label{dudvnew}
				F\left(q,Q[q]\right) =& \int q(u) \left( -\frac{1}{2} u^T D_U' u - \ln q(u) \right) {\rm d}u + \lambda \int q(v)\notag\\& \left( -\frac{1}{2} v^T D_V' v - \ln q(v) \right) {\rm d}v + c',
			\end{align}
			where $D_U' = (I-B)^TW_2^{-1}(I-B)+\lambda A^TW_1^{-1}A$,  $D_V' = (I-A)^TW_1^{-1}(I-A)$, and $c'$ is a constant. 
		\end{corollary}
		\begin{IEEEproof}	
			According to the proof of Theorem \ref{th3}, we get
			\begin{align*}
				d_U\big[Q[q]\big](u) &= - \frac12 u^T(I-B)^TW_2^{-1}(I-B)u+ c\\&=- \frac12 u^TD_Uu+ c_2
			\end{align*}
			and
			\begin{align*}
				&d_V\big[Q[q]\big](v)\\
				=&- \frac12 \big( v^T(W_1^{-1}-A^T W_1^{-1}-W_1^{-1} A+A^TW_1^{-1}A)v\\&\ +{\rm{tr}}(A^TW_1^{-1}AK_U)\big)+ c_4\notag \\
				=&- \frac12 \big( v^TD_V'v+{\rm{tr}}(A^TW_1^{-1}AK_U)\big)+ c_4.
			\end{align*}
			Further, we derive 
			\begin{align*}
				&   \int q(v) d_V\big[Q[q]\big](v) {\rm d}v \\
				=&- \frac12\mathbb {E}_{V} [v^TD_V'v]- \frac12{\rm{tr}}(A^TW_1^{-1}AK_U)+ c_4\notag\\ 
				=\ &  - \frac12\mathbb {E}_{V} [v^TD_V'v]- \frac12{\rm{tr}}(A^TW_1^{-1}A \mathbb{E}_{U} [uu^T])+ c_4\notag\\ 
				=\ &  - \frac12 \int q(v)  v^TD_V'v{\rm d}v- \frac12\int q(u)u^TA^TW_1^{-1}Au{\rm d}u+ c_4.
			\end{align*}
			Substituting the above equations into \eqref{fqQ}, we obtain 
			\begin{align}\label{duu}
				&F(q,Q[q])\notag \\
				= &\int q(u)\left( -\frac{1}{2}u^TD_Uu-\ln q(u)\right) {\rm d}u + \lambda \int q(v) \Big( - \frac12  v^T\notag \\& \ D_V'v-\ln q(v)\Big) {\rm d}v-\frac{\lambda}{2}\int q(u)u^TA^TW_1^{-1}Au{\rm d}u+c'\notag \\
				=&\int q(u)\left( - \frac12 u^T\big(D_U+\lambda A^TW_1^{-1}A\big)u-\ln q(u)\right) {\rm d}u + \lambda\notag \\&\ \int q(v) \left( - \frac12  v^TD_V'v-\ln q(v)\right) {\rm d}v+c'\notag \\	
				=&\int q(u) \left( -\frac{1}{2} u^T D_U' u - \ln q(u) \right) {\rm d}u  \notag \\&+ \lambda \int q(v) \Big( -\frac{1}{2} v^T D_V' v - \ln q(v) \Big) {\rm d}v+c',
			\end{align}
			where $c'=c_2+\lambda c_4$ is a constant.
		\end{IEEEproof}
		
		\begin{remark}\label{rem1}
			It is noteworthy that both Theorem \ref{th3} and Corollary \ref{copo1} provide parallel results that can be applied to subsequent algorithm derivations. In the context of Gaussian vector BC with a covariance constraint studied in this work, the proof of Theorem \ref{th3} specifically utilizes the condition that the covariance sum is constant, leading to relatively simple forms of $D_U$ and $D_V$. Consequently, we primarily adopt the results from Theorem \ref{th3} in our subsequent derivations. However, for scenarios with weaker constraints, such as power constraints, the result in Corollary \ref{copo1} becomes indispensable due to its broader applicability.
		\end{remark}
		
	\subsection{Gaussian Arimoto--Blahut Algorithm with Projection}\label{faal}

Based on the analysis in Section \ref{ASEC}, the form of the optimal $Q$ for a fixed $q$ was derived in Theorem \ref{th1}. Following the steps of the AB algorithm, we now derive the form of the optimal $q$ for a fixed $Q$.
In view of Theorem 2, we consider the problem of maximizing $F(\cdot, Q[q])$ over the pdfs $\tilde{q}(u)$ and $\tilde{q}(v)$ satisfying the covariance constraint $\mathbb{E}_{\tilde{q}(u),\tilde{q}(v)}[ UU^T + VV^T ] = I$, i.e., 
	\begin{equation}
		\begin{aligned}\label{fqQ1}
			&\max_{\tilde{q}(u),\tilde{q}(v)} \int \tilde{q}(u)\left(-\frac{1}{2} u^T D_U u-\ln \tilde{q}(u)\right) {\rm d}u  \\ &\quad\quad\quad+ \lambda \int \tilde{q}(v)\left(-\frac{1}{2} v^T D_V v-\ln \tilde{q}(v)\right) {\rm d}v\\
			& \ \quad  {\rm s.t.} \quad  
			\mathbb{E}_{\tilde{q}(u),\tilde{q}(v)}[UU^T+VV^T] = I.  
		\end{aligned}
	\end{equation} 
	We have the following theorem:
	\begin{theorem}\label{th2}
		Let $\Gamma$ be chosen such that $D_U + \Gamma \succ 0$, $D_V + \Gamma/\lambda \succ 0$, and $(D_U + \Gamma)^{-1} + (D_V + \Gamma/\lambda)^{-1} = I$. Then, the maximizing pdf $\tilde{q}[Q[q]]$ of $F(\cdot,Q[q])$ satisfies
		\begin{equation*}
			\begin{split}
				\tilde{q}\big[Q[q]\big](u)& = g(u; 0, (D_U + \Gamma)^{-1}),    \\
				\tilde{q}\big[Q[q]\big](v)& = g(v; 0, (D_V + \Gamma/\lambda)^{-1}).
			\end{split}
		\end{equation*}
	\end{theorem}
	\begin{IEEEproof}	
		See Appendix \ref{prth3} for details. 
	\end{IEEEproof}	

	Following the framework of the AB algorithm,  the optimization problem \eqref{eqform} can be solved by alternately updating $Q$ and $q$. Specifically, we begin by fixing a joint Gaussian pdf $q$ of $U$ and $V$ whose covariance matrices $K_U$ and $K_V$ satisfy $K_U \succ 0$, $K_V \succ 0$, and $K_U+K_V = I$. After updating $Q$, we obtain $A, W_1$ and $B,W_2$ according to Theorem \ref{th1}, and thus $D_U$ and $D_V$ are determined according to Theorem \ref{th3}. 	  
	Next, we fix the obtained $Q$, which fixes $D_U$ and $D_V$ according to the expression in \eqref{opquv}, and update $q$. According to Theorem 3, the maximizing pdf $\tilde{q}\big[Q[q]\big]$ is jointly Gaussian  when $D_U +\Gamma \succ 0 $ and $D_V  + \Gamma /\lambda \succ 0$ hold, and the covariance matrices $K_U, K_V$ of $U,V$ are given by
	\begin{equation} \label{kukv}
		\begin{split}
			K_U^{-1} = D_U +\Gamma,\quad  K_V^{-1} =  D_V + \Gamma /\lambda,
		\end{split}
	\end{equation}
	respectively. Thus, we can repeat the above process. In essence, the algorithm alternately updates $(D_U, D_V)$ according to Theorems \ref{th1} and \ref{th3} and $(K_U, K_V)$ according to \eqref{kukv}. While it is easy to update $(D_U,D_V)$ given $(K_U,K_V)$, it is not straightforward to update $(K_U,K_V)$ given $(D_U,D_V)$. Indeed,  we need to find a $\Gamma$ that satisfies conditions $D_U +\Gamma \succ 0 $, $D_V  + \Gamma /\lambda \succ 0$ and $(D_U +\Gamma)^{-1}+(D_V + \Gamma /\lambda)^{-1}=I$, which is difficult to solve analytically in the general case (i.e., to obtain a closed-form solution). To overcome this difficulty, we adopt an approximation technique by deriving the equation that $\Gamma$ must satisfy when the algorithm converges and a projection technique to ensure that the covariance constraint is satisfied.
	
		Specifically, based on Theorem \ref{th3}, we obtain
		\begin{align}\label{eq:du}
			D_U	&=(I-B)^TW_2^{-1}(I-B)=(I-B)^TK_U^{-1}\notag\\&=(\Sigma_1(K_U+\Sigma_1)^{-1})^TK_U^{-1}=(K_U\Sigma_1^{-1}(K_U+\Sigma_1))^{-1}\notag\\
			&=(K_U\Sigma_1^{-1}K_U+K_U)^{-1} 
		\end{align}
		and
		\begin{align}\label{dv}
			D_V &= W_1^{-1}(I-A)-A^TW_1^{-1}\notag\\
			&=K_V^{-1}-(I+\Sigma_2)^{-1}(I-A)^{-1}\notag\\
			&=K_V^{-1}-(I+\Sigma_2)^{-1}(I-K_V(I+\Sigma_2)^{-1})^{-1}\notag\\
			&=K_V^{-1}-(I+\Sigma_2)^{-1}((K_U+\Sigma_2)(I+\Sigma_2)^{-1})^{-1}\notag\\
			&=K_V^{-1}-(K_U+\Sigma_2)^{-1}
		\end{align}
		according to the fact that $A=K_V(I+\Sigma_2)^{-1}$, $W_1=(I-A)K_V$, $B=K_U(K_U+\Sigma_1)^{-1}$, and $W_2=(I-B)K_U$.
	
	If the alternating updates converge, then we must have $\Gamma= \lambda(K_U+\Sigma_2)^{-1}$ in the limit based on (\ref{kukv}) and (\ref{dv}). Combining this with \eqref{kukv} and \eqref{eq:du}, we obtain that
	\begin{align}\label{optku}
		K_U =\big((K_U\Sigma_1^{-1}K_U+K_U)^{-1}+\lambda(K_U+\Sigma_2)^{-1}\big)^{-1}
	\end{align}
	in the limit. The fixed-point equation \eqref{optku} suggests a natural iterative procedure for computing the desired $K_U$. Specifically,
	we define $\Pi_{\mathcal{I}}(M )$ as the projection of the PSD matrix $M$ onto the set $\mathcal{I} = \{Y: 0 \preceq Y \preceq I\}$, which is given by $\Pi_{\mathcal{I}}(M)=V\hat D V^T$ with $V D V^T$ being the eigen-decomposition of $M$ and $\hat D_{jj} = \min\{1,D_{jj}\}$. 
	Starting with a $K_U$ satisfying $0 \prec K_U \prec I$, we iteratively compute the right-hand side of \eqref{optku} and project the result onto the set $\mathcal{I}$. This leads to our proposed Gaussian AB algorithm with projection (GAB-P) in the following Algorithm 1.
	\begin{algorithm}[H]
		\caption{GAB-P for
			Gaussian Vector BC with Private Messages}\label{Alg:BA31}
		\begin{algorithmic}[1]
			\REQUIRE $\lambda>1$, $K\succeq 0,\Sigma_1\succ 0,\Sigma_2\succ 0$. 
			\STATE Compute $\hat{{\Sigma}}_{1},\hat{{\Sigma}}_{2}$ based on \eqref{kj}.
			\STATE Initialize $ 0 \prec A_U \prec I$.
			\WHILE {not converge}
			\STATE Update $A_U$:
			\begin{align*}
				A_U&\gets
				\big((A_U\hat{{\Sigma}}_1^{-1}A_U+A_U)^{-1}+\lambda(A_U+\hat{{\Sigma}}_2)^{-1}\big)^{-1}.
			\end{align*}
			\STATE Project $A_U$ onto $\mathcal{I}$:
			\begin{align*}
				A_U \gets \Pi_{\mathcal{I}}(A_U).
			\end{align*} 
			\ENDWHILE
			\STATE Compute $\tilde{K}^{\dag}$ according to (\ref{kn}).
			\ENSURE Covariance matrix $K_U=\tilde{K}^{\dag}A_U{(\tilde{K}^{\dag})}^T$.
		\end{algorithmic}
	\end{algorithm}
	
	It is worth noting that the idea of transforming infinite-dimensional problems into finite-dimensional ones by exploiting the properties of Gaussian distribution also appeared in \cite{uugur2020vector}. Specifically, the authors of \cite{uugur2020vector} considered the vector Gaussian chief executive officer problem under logarithmic loss distortion measure and developed AB-type algorithms to compute its rate-distortion region. Different from \cite{uugur2020vector}, the variables $X$ and $V$ in the capacity region expression (\ref{eq1}) of the Gaussian vector BC are coupled, which creates new challenges to algorithm design. In this paper, we show how to decouple the variables $X$ and $V$ and propose the corresponding Gaussian AB algorithms.

	\subsection{Gaussian Arimoto--Blahut Algorithm with Alternating Updates }\label{BA} 
	
	In Section \ref{faal}, based on the methods in information theory, we derived the form of the optimal solution to the optimization subproblem \eqref{fqQ1} in Theorem \ref{th2} and subsequently developed the heuristic algorithm GAB-P. In this subsection, we present an alternative approach to precisely solve this subproblem, which leads to another AB-type algorithm for the optimization problem \eqref{eq1}.

	Since we have $V\sim \mathcal{N}(0,K_V)$, $U\sim  \mathcal{N}(0,K_U)$, and $K_V+K_U=I$ at optimality, we have the following theorem.
	\begin{theorem}\label{th4}
		For fixed $Q[q]$, the covariance matrices $K_U, K_V$ associated with the optimal solution to the optimization problem \eqref{fqQ1} are optimal for the following optimization problem:
		\begin{equation}
			\begin{aligned}\label{op12}
				&\max_{K_U,K_V\succ 0} -{\rm{tr}}(D_UK_U)-\lambda {\rm{tr}}(D_VK_V)+\ln|K_U|+\lambda \ln|K_V|\\
				& \ \quad  {\rm s.t.} \quad  K_U+K_V=I.
			\end{aligned}
		\end{equation} 
	\end{theorem}
	\begin{IEEEproof}	
		The objective function in the optimization problem \eqref{fqQ1} is
		\begin{align*}
			F(\tilde{q},Q[q])			
			= 	&\int \tilde{q}(u)\left(-\frac{1}{2} u^T D_U u-\ln \tilde{q}(u)\right) {\rm d}u  \\ &+ \lambda \int \tilde{q}(v) \left(-\frac{1}{2} v^T D_V v-\ln \tilde{q}(v)\right) {\rm d}v.
		\end{align*}
		Since $V\sim  \mathcal{N}(0,K_V)$ and $U\sim  \mathcal{N}(0,K_U)$,  we get 
		\begin{align}
			&\int \tilde{q}(u)\left(-\frac{1}{2} u^T D_U u\right) {\rm d}u = -\frac{1}{2}\mathbb {E}_{U} [ {\rm{tr}}(u^TD_Uu)]  \notag\\&= -\frac{1}{2}\mathbb {E}_{U} [ {\rm{tr}}(D_Uuu^T)] = -\frac{1}{2}{\rm{tr}}(D_U\mathbb {E}_{U} [ uu^T])  \notag\\&= -\frac{1}{2}{\rm{tr}} (D_UK_U),\label{12}
		\end{align}
		\vspace{-0.5cm}
		\begin{align}		
			-\int \tilde{q}(u)\ln \tilde{q}(u) {\rm d}u &= h(U)
			\notag\\&= \frac{n}{2}\ln 2\pi+\frac{n}{2}+\frac{1}{2}\ln|K_U|.\label{13}
		\end{align}
		Similarly, we obtain 
		\begin{align}
			\int \tilde{q}(v)\left(-\frac{1}{2} v^T D_V v\right){\rm d}u&= -\frac{1}{2}{\rm{tr}} (D_VK_V), \label{14}\\
			-\int \tilde{q}(v)\ln \tilde{q}(v) {\rm d}v 
			&= \frac{n}{2}\ln 2\pi+\frac{n}{2}+\frac{1}{2}\ln|K_V|\label{15}.
		\end{align}
		Inserting (\ref{12})--(\ref{15}) into the optimization problem (\ref{fqQ1}), we can get the optimization problem (\ref{op12}). 
	\end{IEEEproof}	
	By substituting $K_V=I-K_U$ into the optimization problem (\ref{op12}), we  immediately obtain the following corollary:
	\begin{corollary}\label{co1}
		For fixed $Q[q]$, the covariance matrix $K_U$ associated with the optimal solution to the optimization problem \eqref{fqQ1} is optimal for the following optimization problem: 
		\begin{equation}
			\begin{aligned}\label{optqu}
				\max_{K_U} & -{\rm{tr}}(D_UK_U)-\lambda {\rm{tr}}(D_V(I-K_U))+\ln|K_U|\\ &\quad+\lambda \ln|I-K_U|\\
				{\rm s.t.} & \ 0 \prec K_U\prec I .
			\end{aligned}
		\end{equation}
	\end{corollary}
	
	Let $B=D_U-\lambda D_V=H\tilde{B}H^*$ with $H\tilde{B}H^* $ being the eigen-decomposition of $B$ and $\tilde{B}=\text{diag}(b_1,  b_2,\ldots, b_n)$. We have the following theorem.
	
	\begin{theorem}\label{th5}
		Let $\lambda > 1$ be fixed. The optimal $K_U$ for the optimization problem (\ref{optqu}) is $K_U=HAH^*$, where $A=\text{diag}(a_1,  a_2,\ldots, a_n)$ satisfies 
		\begin{align*}
			a_i=\left\{\begin{matrix}
				\frac{1}{1+\lambda}, & b_i=0,\\
				\frac{(\lambda+1+b_i)-\sqrt{(\lambda+1+b_i)^2-4b_i}}{2b_i}, & b_i\neq 0,
			\end{matrix}\right.		
		\end{align*}
		$\text{for} \ i=1,2,\ldots, n.$
	\end{theorem}
	\begin{IEEEproof}	
		By setting the gradient of the objective function of \eqref{optqu} to zero, we have
		\begin{align*}
			-D_U+K_U^{-1}+\lambda D_V-\lambda (I-K_U)^{-1}=0,
		\end{align*}
		which is equivalent to
		\begin{align}\label{sta}
			K_U^{-1}=D_U-\lambda D_V+\lambda (I-K_U)^{-1}.
		\end{align}
		Given a solution $K_U$ to \eqref{sta}, let $a$ be an eigenvalue and $v$ be a corresponding eigenvector of $K_U$, respectively. Then, we have
		\begin{align}\label{ev}
			\frac{1}{a}v&=(B+\lambda (I-K_U)^{-1} )v=Bv+\frac{\lambda}{1-a}v,
		\end{align}
		or equivalently,
		\begin{align*}
			Bv=\left(\frac{1}{a}-\frac{\lambda}{1-a}\right)v.
		\end{align*}
		It follows that $v$ is also an eigenvector of $B$. Thus, by writing $K_U=HAH^*$ with $A=\text{diag}(a_1,  a_2,\ldots, a_n)$, we get from \eqref{ev} that 
		\begin{align}\label{eqva}
			\frac{1}{a_i}=b_i+\frac{\lambda}{1-a_i}, \quad i=1,2, \ldots, n.
		\end{align}
		Now, let us characterize the solution to \eqref{eqva}.
		\begin{lemma}\label{le1}
			For each $i=1,2,\ldots, n$, the equation \eqref{eqva} has a unique solution in $(0,1)$, which is given by
			\begin{align*}
				a_i=\left\{\begin{matrix}
					\frac{1}{1+\lambda}, & b_i=0,\\
					\frac{(\lambda+1+b_i)-\sqrt{(\lambda+1+b_i)^2-4b_i}}{2b_i}, & b_i\neq 0. 
				\end{matrix}\right.	
			\end{align*}
		\end{lemma}
		\begin{IEEEproof}	
			See Appendix \ref{prba2} for details.
		\end{IEEEproof}	
		Since the optimization problem \eqref{optqu} is convex, Lemma \ref{le1} implies that the unique solution to \eqref{sta} is optimal for \eqref{optqu}. This completes the proof.
	\end{IEEEproof}	
	
	Based on the developments above, we present our proposed Gaussian AB algorithm with alternating updates (GAB-A) in Algorithm \ref{Alg:BA1}.
	\begin{algorithm}[H]
		\caption{GAB-A for
			Gaussian Vector BC with Private Messages}\label{Alg:BA1}
		\begin{algorithmic}[1]
			\REQUIRE $\lambda>1$, $K\succeq 0,\Sigma_1\succ 0,\Sigma_2\succ 0$. 
			\STATE Compute $\hat{{\Sigma}}_{1},\hat{{\Sigma}}_{2}$ based on \eqref{kj}.
			\STATE Initialize $ 0 \prec A_U \prec I$.
			\FOR{$k=1,2,3,\dots$}
			\STATE Update $	D_U,D_V$:
			\begin{align*}
				D_U^{(k)}&\gets(A_U^{(k-1)}\hat{{\Sigma}}_{1}^{-1}A_U^{(k-1)}+A_U^{(k-1)})^{-1},\\
				D_V^{(k)}&\gets(I-A_U^{(k-1)})^{-1}-(A_U^{(k-1)}+\hat{{\Sigma}}_{2})^{-1}.
			\end{align*}
			\STATE Compute the eigen-decomposition $D_U^{(k)} -\lambda D_V^{(k)} = H\text{diag}(b_1,b_2,\ldots,b_n)H^*$.
			\STATE Update $	A_U$:
			\begin{align*}
				A_U^{(k)}\gets H \text{diag}(a_1,  a_2,\ldots, a_n)H^*,
			\end{align*}
			where 
			\begin{align*}
				a_i=\left\{\begin{matrix}
					\frac{1}{1+\lambda}, & b_i=0,\\
					\frac{(\lambda+1+b_i)-\sqrt{(\lambda+1+b_i)^2-4b_i}}{2b_i}, & b_i\neq 0, 
				\end{matrix}\right.					
			\end{align*}
			$\text{for}\  i=1,2,\ldots, n.$
			\ENDFOR
			\STATE Compute $\tilde{K}^{\dag}$ according to (\ref{kn}).
			\ENSURE Covariance matrix $K_U=\tilde{K}^{\dag}A_U{(\tilde{K}^{\dag})}^T$.
		\end{algorithmic}
	\end{algorithm}
	

		It is easy to show that the function values $\{ F(q^{(k)}) \}$ generated by the GAB-A algorithm are monotonically non-decreasing and bounded.
		Indeed, by definition, we have
		\begin{align}\label{fqaqQ}
			\max_{q}F(q)=\max_{q}\max_Q F(q,Q),
		\end{align}
		where $F(q)$ and $F(q,Q)$ are defined in \eqref{fq} and \eqref{fqQ}, respectively.
		When $q$ is fixed, i.e., $K_U$ and $K_V$ are fixed, the optimal $Q$ with the corresponding $D_U$ and $D_V$ for the problem $\max_Q F(q,Q)$ is given by Theorem \ref{th3}. When $Q$ is fixed, i.e., $D_U$ and $D_V$ are fixed, the optimal $K_U$ and $K_V$ are given by Theorem \ref{th5}. This means that both subproblems of the optimization problem $\max_q \max_Q F(q,Q)$ are solved exactly. 			
		Now, let $D_U^{(k)}$ and $D_V^{(k)}$ be the values of $D_U$ and $D_V$ in the $l$-th iteration, respectively; and $K_U^{(k)}$ and $K_V^{(k)}$ be the values of $K_U$ and $K_V$ in the $l$-th iteration, respectively. 
		It follows that 
		\begin{align*}
			F(q^{(k-1)})&\stackrel{(a)}{=}F(q^{(k-1)},Q^{(k-1)})
			\stackrel{(b)}{\leq} F(q^{(k)},Q^{(k-1)}) \\
			&\stackrel{(c)}{\leq} F(q^{(k)},Q^{(k)}) =F(q^{(k)})
		\end{align*}
		for $l=1,2,\ldots$, where $q^{(k)}$ is the joint Gaussian pdf of $U$ and $V$ with covariance matrices $K_U^{(k)}$ and $K_V^{(k)}$, respectively; $Q^{(k)} = Q[q^{(k)}]$ is given by Theorem \ref{th1}. Here, $(a)$ and $(c)$ follow from Theorem \ref{th1} and $(b)$ follows from Theorem \ref{th5}, respectively.
		Moreover, we have  $F(q,Q)\leq F(q)\leq I(X;Y_1)+\lambda I(X;Y_2)$. Therefore, the function value sequence $\{ F(q^{(k)}) \}$ converges.
		

	In addition, for each iteration in Algorithms \ref{Alg:BA31} and \ref{Alg:BA1}, there are $4$ matrix multiplications, $3$ matrix inversions (note that $\hat{\Sigma}_1^{-1}$ needs only to be computed once in the whole algorithm), and $1$ eigen-decomposition. Thus, they have the same computational complexity, which is of order $\mathcal{O} (n^3)$.

		\begin{remark}\label{rem3}
			Observe that $K_U$ is taken to be positive definite rather than positive semidefinite during the course of the GAB-P and GAB-A algorithms, as the density function of the Gaussian distribution is non-degenerate only when the covariance matrix is invertible. This is similar to the classical AB algorithm for discrete distributions, where the $q(x)$ generated in each iteration satisfies $q(x)>0$ due to the exponential update \eqref{ABq}.
		\end{remark}

		\begin{remark}\label{rem2}
			This work focuses on computing the capacity region of Gaussian vector BC subject to a covariance constraint. A closely related optimization problem involves determining the capacity region under a power constraint. Existing approaches indirectly address this non-convex optimization problem by exploiting the duality between BC and MAC \cite{rashid1998transmit, jindal2005sum, yu2006sum, yu2006uplink, yu2007transmitter, zhang2012gaussian}. In contrast, the proposed Gaussian AB framework offers a direct approach to solving this non-convex optimization problem.
			Specifically, we consider the optimization problem 
			\begin{align*}
				\max_{ {X}:{\rm{tr}}(\mathbb {E}[{XX^T}])\leq P  } \lambda I\left(V ; Y_2 \right)+I(X;Y_1|V),
			\end{align*}
			where $V\sim   \mathcal{N}(0, K_V^P )$, $U\sim  \mathcal{N}(0, K_U^P)$ for some $K_U^P, K_V^P \succeq 0$ are independent with $X = V+U$. In the following, we adopt the framework of the GAB-A algorithm. Based on Theorem \ref{th1} and Corollary \ref{copo1}, the optimal $D_U^P$ and $D_V^P$ satisfy $D_U^P=(K_U^P\Sigma_1^{-1}K_U^P+K_U^P)^{-1}+\lambda (K_U^P+\Sigma_2)^{-1}K_V^P(K_U^P+K_V^P+\Sigma_2)^{-1}$ and $D_V^P =(K_V^P)^{-1}-(K_U^P+K_V^P+\Sigma_2)^{-1}$ for a fixed $K_U^P$ and $K_V^P$. Based on Corollary \ref{copo1}, for fixed $D_U^P$ and $D_V^P$, we obtain a convex subproblem with respect to $K_U^P$ and $K_V^P$, i.e., 		
			\begin{equation*}
				\begin{aligned}
					&\max_{K_U^P,K_V^P\succ 0} -{\rm{tr}}(D_U^PK_U^P)-\lambda {\rm{tr}}(D_V^PK_V^P)+\ln|K_U^P|+\lambda \ln|K_V^P|\\
					& \ \quad{\rm s.t.} \ \quad \  {\rm{tr}}(K_U^P+K_V^P)=P.
				\end{aligned}
			\end{equation*}
			by a derivation similar to that of Theorem \ref{th4}. To solve this convex optimization problem, one can use the Lagrange multiplier method to find the optimal $K_U^P$ and $K_V^P$, and the implementation details are omitted here for brevity. As a result, our proposed GAB-A framework can effectively solve capacity region problems with power constraints.
		\end{remark}

	\section{Gaussian Arimoto--Blahut Algorithms with Private and Common Messages}\label{secom}
	In this section, we extend the proposed Gaussian AB algorithms for the Gaussian vector BC with private messages only to the Gaussian vector BC with private and common messages. It is shown in \cite{geng2014capacity} that for $\lambda_0 > \lambda_2 > \lambda_1 > 0$, the capacity region $\hat{\mathcal{C}}$ of the Gaussian vector BC with both private and common messages $p(y_1, y_2|x)$  subject to a covariance matrix constraint $\{X: E[XX^T]\preceq K_C\}$ is characterized by 
	\begin{align}\label{opt2}
		&\max_{(R_0,R_1,R_2) \in \hat{\mathcal{C}} \atop {X}:E[XX^T]\preceq K_C}\lambda_0R_0+\lambda_1R_1+ \lambda_2  R_2 \notag\\	
		= &\max_{K_W,K_V\succeq 0 \atop K_W+K_V\preceq K_C}\lambda_0\min \{I(W;Y_1),I(W;Y_2)\}\notag\\ &\quad\quad\quad\quad\  +\lambda_2I(V;Y_2|W)+ \lambda_1 I(X;Y_1|V,W) \notag\\	
		= &\min_{\alpha \in [0,1]}\max_{K_W,K_V\succeq 0 \atop K_W+K_V\preceq K_C}\alpha\lambda_0I(W;Y_1)+\bar{\alpha}\lambda_0I(W;Y_2)\notag\\ &\quad\quad\quad\quad\quad+\lambda_2I(V;Y_2|W)+ \lambda_1 I(X;Y_1|V,W),
	\end{align}
	where $R_0,R_1,R_2$ are message rates, $\bar{\alpha} = 1-\alpha$, and $W\sim \mathcal{N}(0,K_W)$, $V\sim \mathcal{N}(0,K_V)$, $U\sim \mathcal{N}(0,K_U)$ for some $K_W, K_V, K_U \succeq 0$ are independent with $X = W + V + U$ and $K_C= K_W + K_V + K_U$.
	
	Applying the formula of Gaussian distribution differential entropy \cite{el2011network} to the mutual information expression in \eqref{opt2}, we have
	\begin{align}
		&\alpha\lambda_0I(W;Y_1)+\bar{\alpha}\lambda_0I(W;Y_2)\notag\\ =&\frac{1}{2}\big(\lambda_0(-\alpha\ln|K_C-K_W+\Sigma_1|-\bar{\alpha}\ln|K_C-K_W+\Sigma_2|)\notag\\ &\quad\ + \lambda_0(\alpha\ln|K_C+\Sigma_1|+\bar{\alpha}\ln|K_C+\Sigma_2|)\big),\label{f2} 	
	\end{align}
	\begin{align}
		\lambda_2I(V;Y_2|W) &= \frac{1}{2}\big(\lambda_2\ln|K_C-K_W+\Sigma_2|-\lambda_2\ln|K_C\notag\\ &\quad\ -K_W-K_V+\Sigma_2|\big),\label{vzw}\\
		\lambda_1 I(X;Y_1|V,W) &=\frac{1}{2}\big( \lambda_1\ln|K_C-K_W-K_V+\Sigma_1|\notag\\ &\quad\ -\lambda_1\ln|\Sigma_1|\big)\label{xyvw}
	\end{align}
	for fixed $\alpha \in [0,1]$ and $\lambda_0 > \lambda_2 > \lambda_1$.
	Inserting (\ref{f2})--(\ref{xyvw}) into the objective function of the optimization problem (\ref{opt2}), we obtain 
	\begin{align*}
		&\alpha\lambda_0I(W;Y_1)+\bar{\alpha}\lambda_0I(W;Y_2)+\lambda_2I(V;Y_2|W)\\&+ \lambda_1 I(X;Y_1|V,W) \\
		=& \frac{1}{2}\bigg[\lambda_1\left( \left(\frac{\lambda_2-\lambda_0\bar{\alpha}}{\lambda_1}\right)\ln|K_C-K_W+\Sigma_2|-\frac{\lambda_0\alpha}{\lambda_1}\ln|K_C\right.\\
		&\left.-K_W+\Sigma_1|+\ln|K_C-K_W-K_V+\Sigma_1|-\frac{\lambda_2}{\lambda_1}\ln|K_C\right.\\
		&\left.-K_W-K_V+\Sigma_2|\right)+ \lambda_0(\alpha\ln|K_C+\Sigma_1|+\bar{\alpha}\ln|K_C\\
		&+\Sigma_2|)-\lambda_1\ln|\Sigma_1|\bigg].
	\end{align*}		
	For a fixed $\alpha \in [0,1]$, the quantity $\lambda_0(\alpha\ln|K_C+\Sigma_1|+\bar{\alpha}\ln|K_C+\Sigma_2|)-\lambda_1\ln|\Sigma_1|$ is a constant, and the optimization problem (\ref{opt2}) is equivalent to 
	\begin{align}\label{xy}
		\max_{K_U,K_V\succeq 0\atop K_U + K_V \preceq K_C }&(\lambda_2'-\lambda_0'\bar{\alpha})\ln|K_U + K_V+\Sigma_2| \notag\\
		&-\lambda_0'\alpha\ln|K_U+ K_V+\Sigma_1|+\ln|K_U+\Sigma_1|\notag\\
		&-\lambda_2'\ln|K_U+\Sigma_2|,
	\end{align} 
	where $\lambda_2'=\frac{\lambda_2}{\lambda_1}$,    $\lambda_0'=\frac{\lambda_0}{\lambda_1}$, $\lambda_0'>\lambda_2'>1$, and $K_C-K_W-K_V=K_U$. We may assume that $\lambda_2'-\lambda_0'\bar{\alpha}>0$, since the optimization problem is more tractable in other cases.
	
	Observe that when $K_U + K_V$ is fixed, the optimization problem (\ref{xy}) becomes
	\begin{align}\label{upv}
		\max_{0\preceq K_U\preceq K_U + K_V}\ln|K_U+\Sigma_1|-\lambda_2'\ln|K_U+\Sigma_2|,
	\end{align} 
	which is similar to the optimization problem \eqref{o1} for the case of private messages only and can be solved by the GAB-P or GAB-A algorithm in Section \ref{secmr}.
	Similarly, when $K_U$ is fixed, the optimization problem (\ref{xy}) becomes
	\begin{align}
		\max_{K_U\preceq K_U + K_V\preceq K_C }&(\lambda_2'-\lambda_0'\bar{\alpha})\ln|K_U + K_V+\Sigma_2| \notag\\
		&-\lambda_0'\alpha\ln|K_U+ K_V+\Sigma_1|,\notag
	\end{align} 
	or equivalently,
	\begin{align}\label{xfix}
		\max_{0\preceq K_V\preceq K_C-K_U }&(\lambda_2'-\lambda_0'\bar{\alpha})\ln|K_V+(K_U+\Sigma_2)|\notag\\
		&-\lambda_0'\alpha\ln|K_V+(K_U+\Sigma_1)|,
	\end{align}
	which can also be solved by the GAB-P and GAB-A algorithms with minor modifications.
	
	From the above, we see that the optimization problem \eqref{xy} can be addressed by alternately updating the variables $K_U$ and $K_U + K_V$  via solving \eqref{upv} and \eqref{xfix}, respectively.
	Unfortunately, for the optimization subproblem \eqref{upv}, the corresponding algorithm is sensitive to the initial value of $K_U + K_V$. Thus, we consider fixing $K_V$ and updating $K_U$ by solving the optimization subproblem 
	\begin{align}\label{vop}
		\max_{0\preceq K_U\preceq K_C-K_V}&(\lambda_2'-\lambda_0'\bar{\alpha})\ln|K_U+(K_V+\Sigma_2)|\notag\\
		&-\lambda_0'\alpha\ln|K_U+(K_V+\Sigma_1)|+\ln|K_U+\Sigma_1|\notag\\
		&-\lambda_2'\ln|K_U+\Sigma_2|,
	\end{align} 
	which can be done using the techniques developed in Section \ref{secmr}.
	In sum, we propose to solve the optimization problem \eqref{xy} by alternately updating $K_U$ and $K_V$ via solving \eqref{upv} and \eqref{vop}, respectively. To implement this approach, we can extend, e.g., the GAB-P algorithm in Section \ref{faal}, leading to the extended GAB-P algorithm (EGAB-P) in Algorithm \ref{Alg:BA214}. The detailed derivation of Algorithm \ref{Alg:BA214} can be found in Appendix \ref{prcom}.
	
	\begin{algorithm}[H]
		\caption{EGAB-P for
			Gaussian Vector BC with Both Private and Common Messages}\label{Alg:BA214}
		\begin{algorithmic}[1]
			\REQUIRE $\lambda_0/\lambda_1>\lambda_2/\lambda_1>1$, $K\succeq 0,\Sigma_1\succ 0,\Sigma_2\succ 0$, $\alpha \in [0,1]$. 
			\STATE Initialize $ 0 \prec K_U \prec K_C$.
			\WHILE {not converge}			
			\STATE Let $K= K_C-K_U$, $N_1=K_U+\Sigma_2$, and $N_2=K_U+\Sigma_1$ and compute $\hat{N}_{1},\hat{N}_{2}$ based on \eqref{kj}.
			\STATE Initialize $ 0 \prec B_V \prec I$.
			\WHILE {not converge}
			\STATE Update $B_V$:
			\begin{align*}
				B_V&\gets
				\left((B_V\hat{N}_1^{-1}B_V+B_V)^{-1}+\frac{\lambda_0\alpha}{\lambda_2-\lambda_0\bar{\alpha}}\right.\\&\quad\quad\left.(B_V+\hat{N}_2)^{-1}\right)^{-1}.
			\end{align*}
			\STATE Project $B_V$ onto $\mathcal{I}$: $B_V \gets \Pi_{\mathcal{I}}(B_V)$.
			\ENDWHILE
			\STATE Compute covariance matrix $K_V=\tilde{K}^{-1}B_V{(\tilde{K}^{-1})}^T$ according to (\ref{kn}).
			\STATE Let $K= K_C-K_V$, $M_1=K_V+\Sigma_2$, and $M_2=K_V+\Sigma_1$. Compute $ \hat{M}_{1},\hat{M}_{2},\hat{{\Sigma}}_{1},\hat{{\Sigma}}_{2}$ based on \eqref{kuk} and \eqref{kj}. Compute $\tilde{K}_V$ based on \eqref{kuk} and denote the non-zero submatrix in the upper left corner of $\tilde{K}_V$ as $B_V'$.
			\STATE Initialize $ 0 \prec A_U \prec I$.
			\WHILE {not converge}
			\STATE Update $A_U$:
			\begin{align*}
				A_U&\gets \left((A_U\hat{{\Sigma}}_{1}^{-1}A_U+A_U)^{-1}+\frac{\lambda_2}{\lambda_1}(A_U+\hat{{\Sigma}}_{2})^{-1}\right.\\
				&\quad\quad \ \left.B_V'(A_U+\hat{M}_{1})^{-1} +\frac{\lambda_0}{\lambda_1}\left(\alpha(A_U+\hat{M}_{2})^{-1}\right.\right.\\
				&\quad\quad \ \left.\left.+{\bar{\alpha}}(A_U+\hat{M}_{1})\right)\right)^{-1}.
			\end{align*}
			\STATE Project $A_U$ onto $\mathcal{I}$: $A_U \gets \Pi_{\mathcal{I}}(A_U)$.
			\ENDWHILE
			\STATE Compute covariance matrix $K_U=\tilde{K}^{\dag}A_U{(\tilde{K}^{\dag})}^T$ according to (\ref{kn}).
			\ENDWHILE			
			\ENSURE Covariance matrices $K_U$ and $K_V$.
		\end{algorithmic}
	\end{algorithm}
	
	\section{Numerical Simulations}\label{secop}
	In this section, we evaluate the performance of the proposed Gaussian AB algorithms by numerical simulations.
	\subsection{The Case with Private Messages}\label{numpri}		
	In this subsection, we demonstrate the performance of the GAB-P and GAB-A algorithms for computing the capacity region of the Gaussian vector BC with private messages. We consider the cases where $n=2$ and $n$ is large, where $n$ denotes the dimension of the matrix in the optimization problem.
	
	Let $K_U^*:= (\Sigma_2-\lambda\Sigma_1)/(\lambda-1)$ be the point at which the gradient of the objective function of \eqref{o1} is zero. In terms of the relationship between $K_U^*$ and the feasible set of the optimization problem \eqref{eq1}, we consider four cases:
	1) $K_U^* \in \mathbb{S}_+\cap \mathbb{S}_K$, 2) $K_U^* \in (\mathbb{S}_+)^c \cap \mathbb{S}_K$, 3) $K_U^* \in \mathbb{S}_+ \cap (\mathbb{S}_K)^c$, 4) $K_U^* \in  (\mathbb{S}_+ \cup \mathbb{S}_K)^c$. In the first case, $K_U^*$ is feasible for \eqref{eq1}. In all the remaining cases, $K_U^*$ is infeasible. In the following, we construct four examples with $n=2$ corresponding to the four cases and compare the solutions obtained by an exhaustive search algorithm (denoted by $K_U^E$), by the proposed GAB-P algorithm (denoted by $K_U^P$), and by the proposed GAB-A algorithm (denoted by $K_U^A$). We set $\lambda=2>1$ and denote the values of the objective function in (\ref{eq1}) at $K_U^{E}$, $K_U^{P}$, and $K_U^{A}$ as $f^{E}$, $f^{P}$, and $f^{A}$, respectively.
	
	1) $K_U^* \in \mathbb{S}_+ \cap \mathbb{S}_K$: 
	We take
	\begin{align*}
		\begin{aligned}
			\Sigma_1=\begin{bmatrix}
				1 & 0 \\
				0 & 1 
			\end{bmatrix}
		\end{aligned}
		,
		\begin{aligned}
			\Sigma_2=\begin{bmatrix}
				3 & 1 \\
				1 & 4
			\end{bmatrix}
		\end{aligned}
		,
		\begin{aligned}
			K=\begin{bmatrix}
				2 & 2 \\
				2 & 4
			\end{bmatrix}
		\end{aligned}
		,
		\begin{aligned}
			K_U^*=\begin{bmatrix}
				1 & 1 \\
				1 & 2
			\end{bmatrix}
		\end{aligned}
		.
	\end{align*}
	In this case, $K_U^*$ is optimal for \eqref{eq1}, and the solutions obtained by our GAB-P and GAB-A algorithms are exactly $K_U^*$, i.e., $K_U^P = K_U^A = K_U^*$.
	
	2)  $K_U^* \in (\mathbb{S}_+)^c \cap \mathbb{S}_K$: 
	We take
	\begin{align*}
		\begin{aligned}
			\Sigma_1=\begin{bmatrix}
				1 & 0 \\
				0 & 1 \\
			\end{bmatrix}
		\end{aligned}
		,
		\begin{aligned}
			\Sigma_2=\begin{bmatrix}
				3 & 2 \\
				2 & 4
			\end{bmatrix}
		\end{aligned}
		,
		\begin{aligned}
			K=\begin{bmatrix}
				2 & 2 \\
				2 & 4
			\end{bmatrix}
		\end{aligned}
		,
		\begin{aligned}
			K_U^*=\begin{bmatrix}
				1 & 2 \\
				2 & 2 
			\end{bmatrix}
		\end{aligned}.
	\end{align*}
	In this case, we have 
	\begin{align*}
		\begin{aligned}
			K_U^{E}=\begin{bmatrix}
				1.3520   & 1.7305 \\
				1.7305 & 2.2150 
			\end{bmatrix}
		\end{aligned}
		,
		\begin{aligned}
			K_U^{P}= K_U^{A}=\begin{bmatrix}			   
				1.3489 & 1.7276 \\
				1.7276  & 2.2127
			\end{bmatrix}
		\end{aligned},
	\end{align*}
	where $\left \| K_U^{P}- K_U^{E} \right \|_{2}= 2.1356\times 10^{-4}$ and $\left \|K_U^{P}- K_U^{A}\right \|_{2}=7.8773\times 10^{-9}$. In addition, we have $f^{P}-f^{E}=1.4449\times 10^{-6}>0$ and $f^{P}-f^{A}=-1.6387\times 10^{-13}<0$.
	
	3) $K_U^* \in \mathbb{S}_+ \cap (\mathbb{S}_K)^c$: 
	We take
	\begin{align*}
		\begin{aligned}
			\Sigma_1=\begin{bmatrix}
				1 & 0 \\
				0 & 1 \\
			\end{bmatrix}
		\end{aligned}
		,
		\begin{aligned}
			\Sigma_2=\begin{bmatrix}
				5 & 2 \\
				2 & 4
			\end{bmatrix}
		\end{aligned}
		,
		\begin{aligned}
			K=\begin{bmatrix}
				2 & 2 \\
				2 & 4
			\end{bmatrix}
		\end{aligned}
		,
		\begin{aligned}
			K_U^*=\begin{bmatrix}
				3 & 2 \\
				2 & 2 
			\end{bmatrix}
		\end{aligned}.
	\end{align*}
	In this case, we have
	\begin{align*}
		\begin{aligned}
			K_U^{E}=
			K_U^{P}= K_U^{A}=\begin{bmatrix}			   
				1.9170 & 1.5760 \\
				1.5760  & 1.8340
			\end{bmatrix}
		\end{aligned},
	\end{align*}
	where $\left \|K_U^{P}- K_U^{E}\right \|_2= 5.0080\times 10^{-6}$ and $\left \|K_U^{P}- K_U^{A}\right \|_2=4.0490\times 10^{-5}$. In addition, we have $f^{P}-f^{E}=9.9210\times 10^{-13}>0$ and $f^{P}-f^{A}=1.9991\times 10^{-6}>0$.
	
	4) $K_U^* \in (\mathbb{S}_+ \cup \mathbb{S}_K)^c$: 
	We take
	\begin{align*}
		\begin{aligned}
			\Sigma_1=\begin{bmatrix}
				1 & 0 \\
				0 & 1 \\
			\end{bmatrix}
		\end{aligned}
		,
		\begin{aligned}
			\Sigma_2=\begin{bmatrix}
				3 & 2 \\
				2 & 4
			\end{bmatrix}
		\end{aligned}
		,
		\begin{aligned}
			K=\begin{bmatrix}
				1 & 1 \\
				1 & 4
			\end{bmatrix}
		\end{aligned}
		,
		\begin{aligned}
			K_U^*=\begin{bmatrix}
				1 & 2 \\
				2 & 2 
			\end{bmatrix}
		\end{aligned}.
	\end{align*}
	In this case, we have
	\begin{align*}
		\begin{aligned}
			K_U^{E}=\begin{bmatrix}
				0.9530 & 1.3194 \\
				1.3194 & 1.8270 
			\end{bmatrix}
		\end{aligned}
		,
		\begin{aligned}
			K_U^{P}= K_U^{A}=\begin{bmatrix}			   
				0.9536 & 1.3179 \\
				1.3179  & 1.8215
			\end{bmatrix}
		\end{aligned},
	\end{align*}
	where $\left \|K_U^{P}- K_U^{E}\right \|_2= 0.0059$ and $\left \|K_U^{P}- K_U^{A}\right \|_2=5.0681\times 10^{-5}$. In addition, we have $f^{P}-f^{E}=4.2013\times 10^{-5}>0$ and $f^{P}-f^{A}=1.9986\times 10^{-6}>0$.
	
	All the above results demonstrate the effectiveness of our proposed GAB-P and GAB-A algorithms.
	
	\begin{figure}[h]
		\centering
		\includegraphics[width=2.5in]{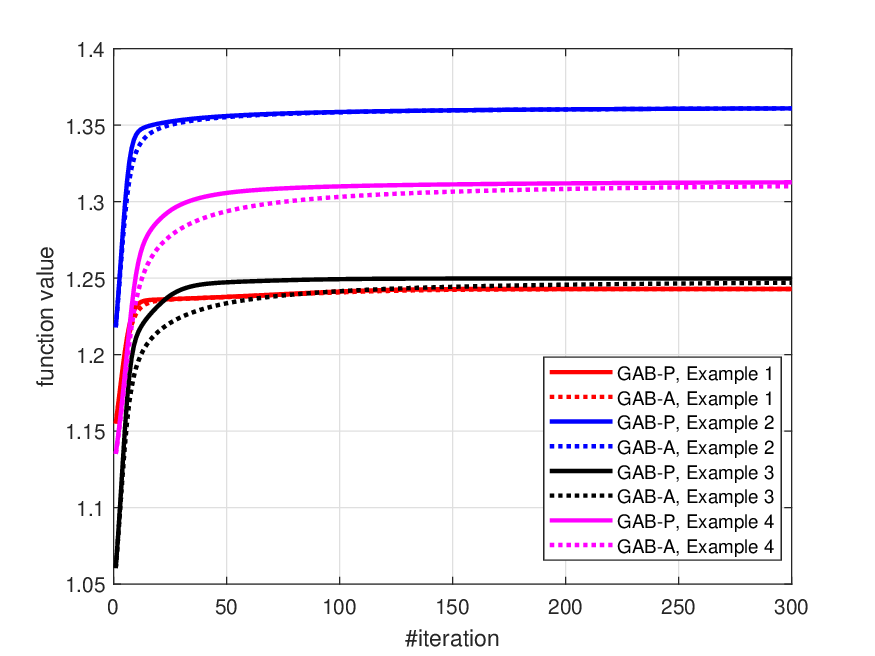}
		\caption{Function values versus number of iterations for the four examples above.}
		\label{fig4}
	\end{figure}			
	Fig. \ref{fig4} depicts the objective value of \eqref{eq1} versus the number of iterations of the proposed GAB-P and GAB-A algorithms for the four examples above. It is observed that both algorithms converge quickly, with the GAB-P algorithm converging even more rapidly.

\begin{figure}[h]
			\centering
			\includegraphics[width=2.5in]{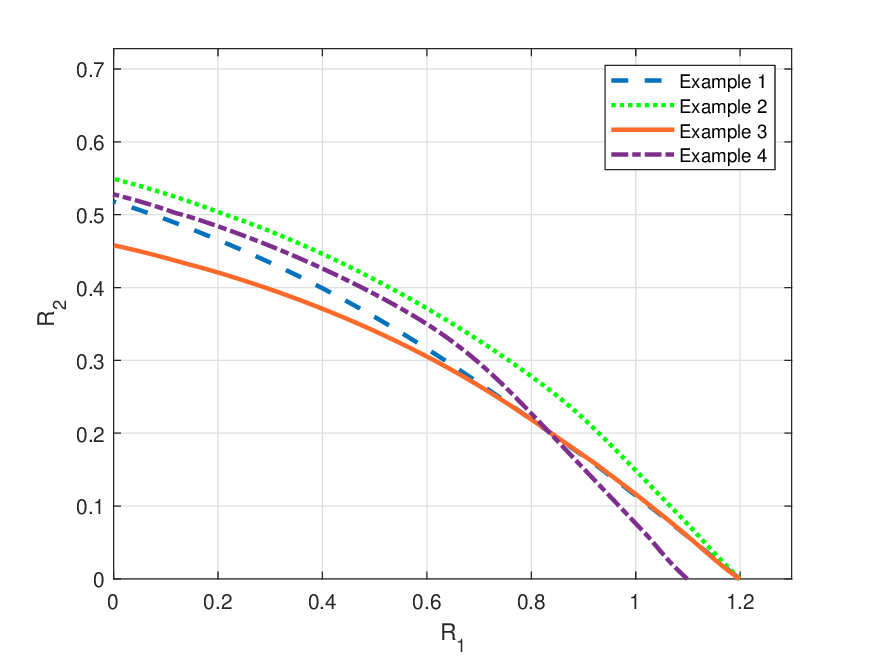}
			\caption{Capacity regions for the four examples above.}
			\label{figc}
		\end{figure}

		Then, we obtain the capacity regions of the Gaussian vector BCs corresponding to the above four examples in Fig. \ref{figc} by traversing $\lambda$. Specifically, we solve the optimization problem \eqref{eq1} with $R_1+ \lambda R_2$ and $R_2+ \lambda R_1$ as objective functions for $\lambda >1$, respectively. In the latter case, the roles of $R_1$ and $R_2$ are interchanged in the proposed Gaussian AB algorithms. A comparison of the channel models in Examples 2 and 3 reveals that the noise covariance matrix of Example 3 is greater than that of Example 2 in the positive semidefinite ordering. Consequently, the capacity region of Example 3 is contained within that of Example 2, as presented in Fig. \ref{figc}.

	For larger $n$, we compare our proposed GAB-P and  GAB-A algorithms with the DCProx algorithm in \cite{yao2023globally}.
	\begin{figure}[h]
		\centering
		\includegraphics[width=2.5in]{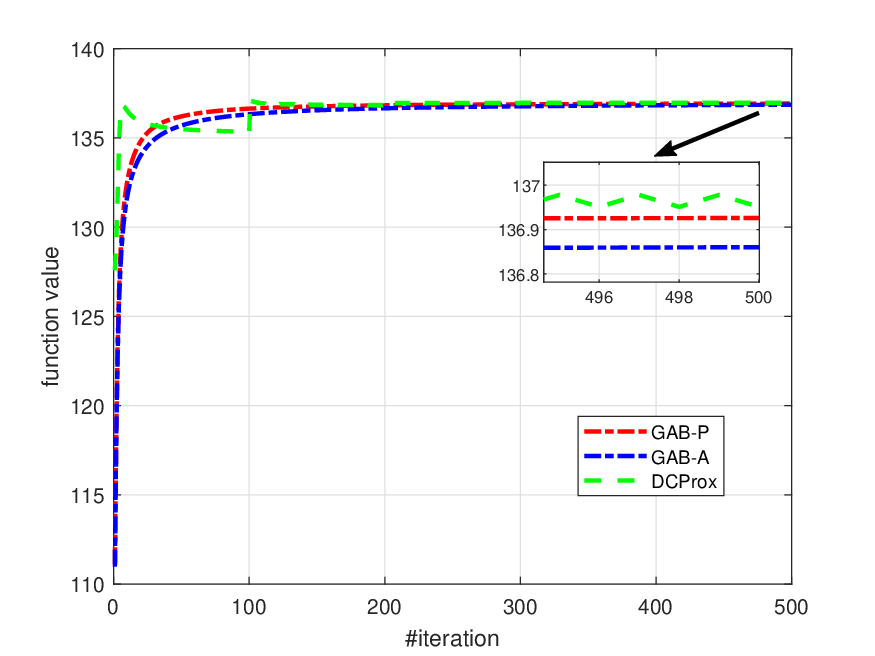}
		\caption{Function values of different algorithms versus number of iterations.}
		\label{fig1}
	\end{figure}
	Fig. \ref{fig1} depicts the objective value of (\ref{eq1}) versus the number of iterations with $n=100$. It is observed that our proposed GAB-P and GAB-A algorithms, as well as the DCProx algorithm with appropriate parameters, all exhibit satisfactory convergence performance. However, the DCProx algorithm suffers from fluctuations and the corresponding solution falls outside of the feasible set. In contrast, our proposed algorithms do not require parameter tuning and guarantee that each iterate is feasible.
	
	\begin{figure}[h]
		\centering
		\includegraphics[width=2.5in]{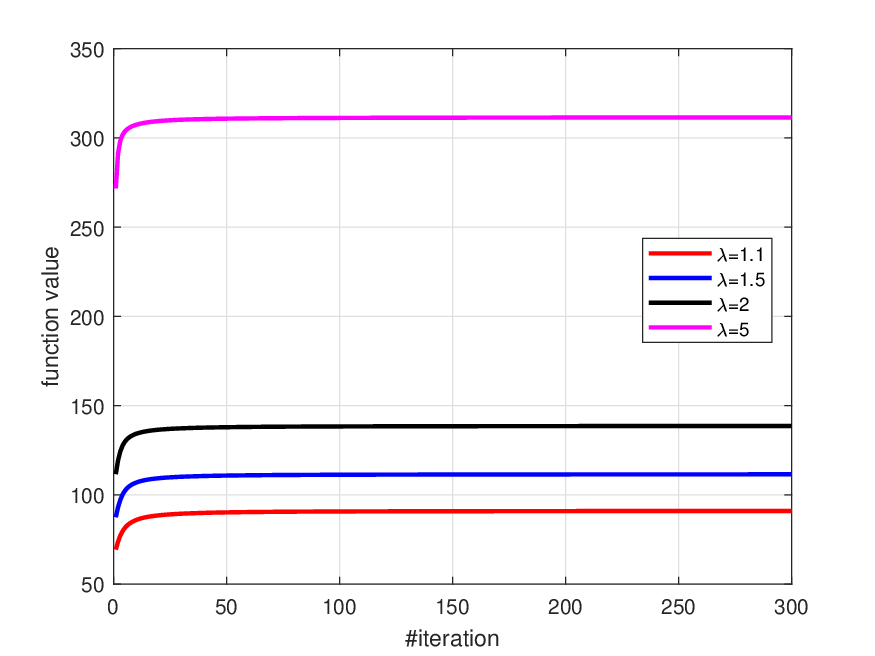}
		\caption{Function values versus number of iterations of the GAB-P algorithm with different $\lambda$.}
		\label{fig2}
	\end{figure}
	Next, we present the performance of our proposed GAB-P algorithm under varying values of $\lambda$ with $n=100$ in  Fig. \ref{fig2}.	As observed in Fig. \ref{fig2}, the objective value of \eqref{eq1} increases as $\lambda$ increases, and our algorithm demonstrates stable convergence performance across different values of $\lambda$.				
	
\begin{figure}[h]
		\centering
		\includegraphics[width=2.5in]{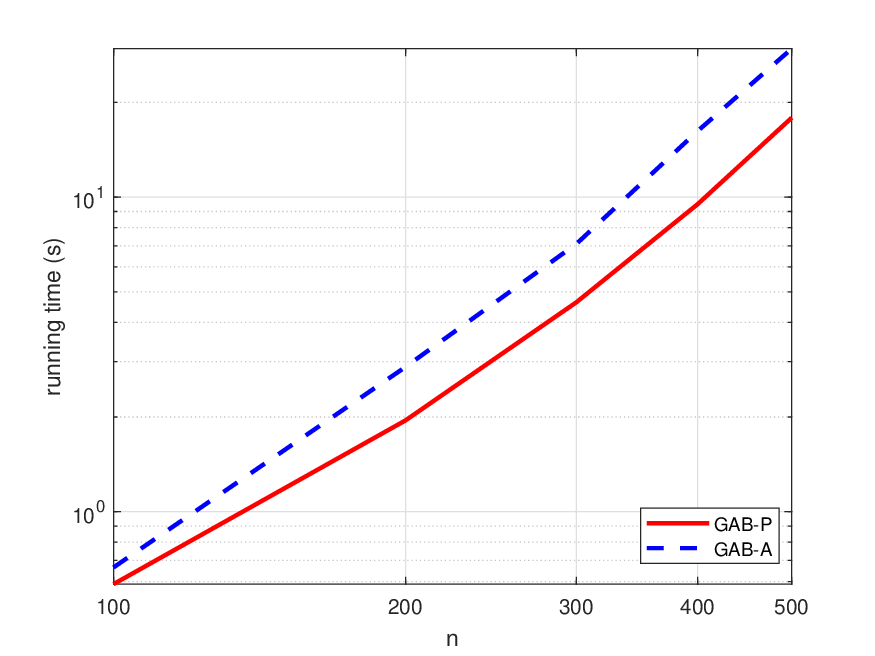}
		\caption{The average running time of the GAB-P and GAB-A algorithms versus the matrix dimension $n$.}
		\label{fig3}
	\end{figure}

	Furthermore, we compare the average running time of the GAB-P and GAB-A algorithms versus the matrix dimension $n$ in Fig. \ref{fig3}, where $n$ varies from $100$ to $500$ over $20$ random tests. 
	In Fig. \ref{fig3} plotted on a log-log scale, it is observed that the running time of the algorithms increases as the problem dimension grows. The average running time of the GAB-P algorithm is less than that of the GAB-A algorithm since the former takes fewer iterations to converge.

	\subsection{The Case with Private and Common Messages}\label{numpricom}
	
\begin{figure}[h]
		\centering
		\includegraphics[width=2.5in]{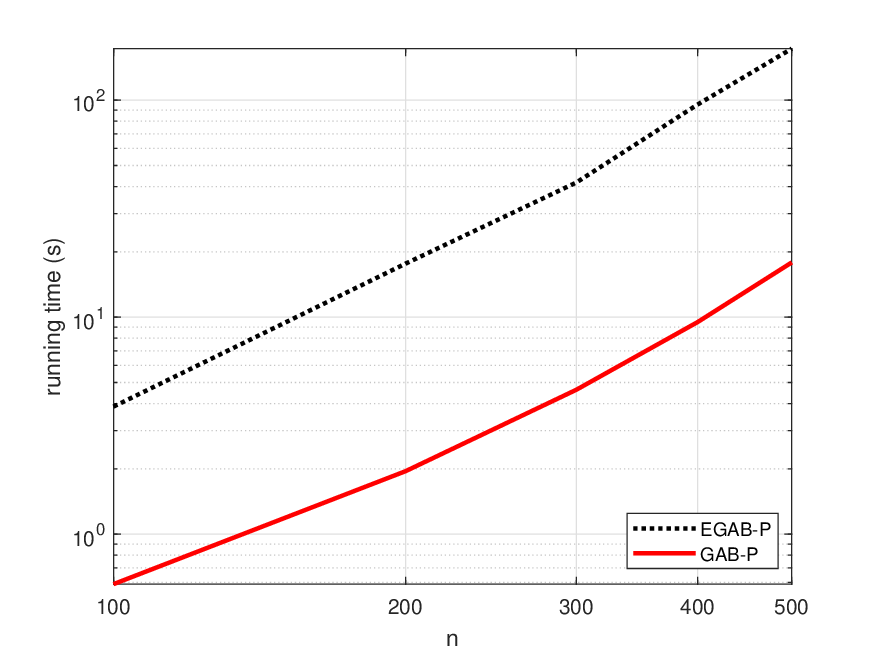}
		\caption{The average running time of the EGAB-P and GAB-P algorithms versus the matrix dimension $n$.}
		\label{fig5}
	\end{figure}

	In this subsection, we evaluate the performance of the EGAB-P algorithm for computing the capacity region of the Gaussian vector BC with private and common messages.		
	We present the average running time of the EGAB-P algorithm versus $n$ in Fig. \ref{fig5} with $n$ varying from $100$ to $500$. 
	The parameters in Algorithm \ref{Alg:BA214} are set to $\lambda_0=1.2>\lambda_2=1.1>\lambda_1=1$ and $\alpha=0.5$.				
	
	By comparing the running time of GAB-P and EGAB-P in Fig. \ref{fig5}, it is seen that the latter incurs a longer running time than the former due to the increased number of optimization variables introduced by common messages.
	
	\section{Conclusions}\label{secco}
	In this paper, we developed discretization-free and parameter-free Gaussian AB algorithms to calculate the capacity region of the Gaussian vector BC. 
	Within the framework of the AB algorithm, we transformed the original optimization problems, which involve pdfs and are infinite-dimensional, into optimization problems that involve covariance matrices and are finite-dimensional. This is achieved by leveraging the property of the Gaussian distribution. Then, we developed two algorithms, namely GAB-P and GAB-A, to solve the finite-dimensional optimization problems.
	Moreover, we developed an extension of the GAB-P algorithm, called the EGAB-P algorithm, to compute the capacity region of the Gaussian vector BC with both private and common messages. 
	We conducted numerical simulations to verify the effectiveness of our proposed algorithms. 
	One interesting future direction is to extend our theoretical framework to solve optimization problems involving other distribution families that can be represented by finite parameters.

	\appendices
	
	\section{Proof of Theorem \ref{th2}} \label{prth3}
	
	Inspired by the proof of \cite[Theorem 2.50]{yeung2008information}, for any joint pdf $\tilde{q}$ satisfying the constraint $\mathbb{E}_{\tilde{q}(u,v)}[UU^T+VV^T] = I$, we have
	\begin{align*}
		& f(\tilde{q}\big[Q[q]\big],Q[q]) - f(\tilde{q},Q[q])  \\
		= &\mathbb E_{\tilde{q}\big[Q[q]\big](u)}   \left[-\frac{1}{2} U^T D_U U-\ln \tilde{q}\big[Q[q]\big](U)\right]+ \lambda \mathbb E_{\tilde{q}\big[Q[q]\big](v)} \\&\left[-\frac{1}{2} V^T D_V V-\ln \tilde{q}\big[Q[q]\big](V)\right]- \mathbb E_{\tilde{q}(u)}\left[-\frac{1}{2} U^T D_U U\right.\\&\left.-\ln \tilde{q}(U)\right]  			
		- \lambda\mathbb E_{\tilde{q}(v)} \left[-\frac{1}{2} V^T D_V V-\ln \tilde{q}(V)\right]\\
		= &\mathbb E_{\tilde{q}\big[Q[q]\big](u)}   \left[-\frac{1}{2} U^T D_U U-\left(-\frac{1}{2} U^T (D_U + \Gamma) U \right)\right]\\&+ \lambda \mathbb E_{\tilde{q}\big[Q[q]\big](v)}\bigg[-\frac{1}{2} V^T D_V V- \bigg(-\frac{1}{2} V^T (D_V + \Gamma/\lambda) \\	
		&V\bigg)\bigg]- \mathbb E_{\tilde{q}(u)}\left[-\frac{1}{2} U^T D_U U-\ln \tilde{q}(U)\right]  
		- \lambda \mathbb E_{\tilde{q}(v)}\bigg[-\frac{1}{2}  \\&V^TD_V V-\ln \tilde{q}(V)\bigg]+c\\
		=&\frac{1}{2}\mathbb E_{\tilde{q}\big[Q[q]\big](u)} \left[U^T\Gamma U \right] + \frac{\lambda}{2} \mathbb E_{\tilde{q}\big[Q[q]\big](v)}\left[V^T\Gamma V /\lambda\right]- \mathbb E_{\tilde{q}(u)}\\&\left[-\frac{1}{2} U^T D_U U-\ln \tilde{q}(U)\right]- \lambda \mathbb E_{\tilde{q}(v)} \bigg[-\frac{1}{2} V^T D_V V-\ln \\&\ \tilde{q}(V)\bigg]+c\\
		\stackrel{(a)}{=}&-\mathbb E_{\tilde{q}(u)}\left[-\frac{1}{2} U^T (D_U+\Gamma)U\right]- \lambda \mathbb E_{\tilde{q}(v)} \bigg[-\frac{1}{2} V^T (D_V\\&+\Gamma/\lambda) V\bigg] + \mathbb E_{\tilde{q}(u)}\left[\ln \tilde{q}(U)\right]+\lambda \mathbb E_{\tilde{q}(v)} \left[\ln \tilde{q}(V)\right]+c\\
		=&-\mathbb E_{\tilde{q}(u)}\left[\ln \tilde{q}\big[Q[q]\big](U)\right] -\lambda \mathbb E_{\tilde{q}(v)} \left[\ln \tilde{q}\big[Q[q]\big](V)\right] \\&+\mathbb E_{\tilde{q}(u)}[\ln \tilde{q}(U)]+ \lambda \mathbb E_{\tilde{q}(v)} \left[\ln \tilde{q}(V)\right]\\
		= &\mathbb E_{\tilde{q}(u)}\left[\ln \frac{\tilde{q}(U)}{\tilde{q}\big[Q[q]\big](U)} \right]+\lambda \mathbb E_{\tilde{q}(v)} \left[ \ln \frac{\tilde{q}(V)}{\tilde{q}\big[Q[q]\big](V)} \right] \\
		= & D\big( \tilde{q}(u) \, \| \, \tilde{q}\big[Q[q]\big](u) \big) +\lambda  D\big( \tilde{q}(v) \, \| \, \tilde{q}\big[Q[q]\big](v) \big)\\
		\geq& 0,								
	\end{align*}
	where (a) holds by
	\begin{align*}
		&\mathbb E_{\tilde{q}(u)} [U^T\Gamma U ]+\lambda\mathbb E_{\tilde{q}(v)} [ V^T\Gamma V/\lambda]
		= \mathbb E_{\tilde{q}(u)} [\mathrm{tr}(\Gamma UU^T) ]\\&+\mathbb E_{\tilde{q}(v)} [ \mathrm{tr}(\Gamma VV^T)]
		=  \mathrm{tr}\left(\Gamma (E_{\tilde{q}(u)}[UU^T]+E_{\tilde{q}(v)}[VV^T])\right)
		\\&= \mathrm{tr}(\Gamma)\mathbb= E_{\tilde{q}\big[Q[q]\big](u)} [U^T\Gamma U ]+\lambda\mathbb E_{\tilde{q}\big[Q[q]\big](v)} [ V^T\Gamma V/\lambda].
	\end{align*}			
	This completes the proof.

	\section{Proof of Lemma \ref{le1}} \label{prba2}	
	For convenience, we omit the subscript $i$ in (\ref{eqva}). We prove that the equation (\ref{eqva}) admits two solutions, one of which satisfies $0 \leq a\leq 1$ and the other satisfies $a>1$ or $a<0$.
	
	For the eigenvalue $a$ of $K_U$ and the corresponding eigenvalue $b$ of $B$, according to (\ref{eqva}), we get
	\begin{align}\label{2equ}
		ba^2-(\lambda+1+b)a+1=0.
	\end{align}
	Consider the quadratic equation $	bx^2-(\lambda+1+b)x+1=0$, where $b\not=0$. Its discriminant satisfies $(\lambda+1+b)^2-4b=(\lambda-1+b)^2+4\lambda\geq 0$ since $\lambda>1$. Thus, the equation admits the solutions $$x_1=\frac{(\lambda+1+b)-\sqrt{(\lambda+1+b)^2-4b}}{2b}$$ 
	and $$x_2=\frac{(\lambda+1+b)+\sqrt{(\lambda+1+b)^2-4b}}{2b}.$$
	In the following, we show that 
	\begin{align*}
		\begin{aligned}
			0< x_1< 1
		\end{aligned}
		,\quad
		\begin{aligned}
			\left\{\begin{matrix}
				x_2>1, & b>0,\\
				x_2<0, & b<0.
			\end{matrix}\right.		
		\end{aligned}
	\end{align*}
	
	(1) $ 0< x_1< 1 $.
	
	For $ b>0 $, we have $\sqrt{(\lambda+1+b)^2-4b}< \lambda+1+b$, which means that $x_1>0$. On the other hand, $x_1< 1$ is equivalent to $(\lambda+1+b)-\sqrt{(\lambda+1+b)^2-4b} < 2b$, i.e., $(\lambda+1+b)-2b <\sqrt{(\lambda+1+b)^2-4b}$. Obviously, this last inequality holds when $(\lambda+1+b)-2b=\lambda+1-b\le0$. In addition, when $\lambda+1-b>0$, we have $\big((\lambda+1+b)-2b\big)^2 < (\lambda+1+b)^2-4b$ because
	$\big((\lambda+1+b)-2b\big)^2=(\lambda+1+b)^2-4b-4b\lambda,\ \lambda>1$, and $b>0$.
	
	For $ b<0 $, we have $\sqrt{(\lambda+1+b)^2-4b}>|\lambda+1+b| >\lambda+1+b$, which implies that $x_1 > 0$. On the other hand, $x_1< 1$ is equivalent to
	$(\lambda+1+b)-\sqrt{(\lambda+1+b)^2-4b} > 2b$. This last inequality can be shown to hold by following a similar argument as above and noting that $-4b\lambda>0$ due to $\lambda>1$ and $b<0$.
	
	(2) $x_2> 1 $ for $b>0$; $x_2<0 $ for $b<0$.
	
	For $ b>0 $, $x_2> 1$ is equivalent to 
	$(\lambda+1+b)+\sqrt{(\lambda+1+b)^2-4b} > 2b$, i.e., 
	$(\lambda+1+b)-2b >-\sqrt{(\lambda+1+b)^2-4b}$. Obviously, this last inequality holds when $\lambda+1-b\geq 0$. On the other hand, when $\lambda + 1 - b < 0$, we get $\big((\lambda+1+b)-2b\big)^2 < (\lambda+1+b)^2-4b$ because $-4b\lambda<0$ due to $\lambda>1$ and $b>0$.
	
	For $b<0$, $x_2<0$ is equivalent to
	$(\lambda+1+b)+\sqrt{(\lambda+1+b)^2-4b} > 0$, i.e.,
	$(\lambda+1+b) >-\sqrt{(\lambda+1+b)^2-4b}$. Obviously, this last inequality holds when $\lambda+1+b\geq 0$. When $\lambda + 1 + b < 0$, we get $(\lambda+1+b)^2 < (\lambda+1+b)^2-4b$. This completes the proof.

	\section{Algorithm for solving optimization problem (\ref{vop})} \label{prcom}	
	According to the  mutual information expression in (\ref{opt2}), we formulate the objective function as follows:
	\begin{align*}
		&F_C(q,Q)\\
		=&\alpha\lambda_0I(W;Y_1)+\bar{\alpha}\lambda_0I(W;Y_2)+\lambda_2I(V;Y_2|W)\\&+ \lambda_1 I(X;Y_1|V,W) \\
		=&  \alpha\lambda_0 \big(h(W)-h(W| Y_1)\big) +  \bar{\alpha}\lambda_0 \big(h(W)-h(W|Y_2)\big)+\lambda_2\big(h\\&(V|W)-h(V|W,Y_2)\big)  +\lambda_1(h(X|V,W)-h(X|Y_1,V,W)) \\
		=&  \alpha\lambda_0 \big(h(W)-h(W| Y_1)\big) +  \bar{\alpha}\lambda_0 \big(h(W)-h(W|Y_2)\big)+\lambda_2\\&\big(h(V)-h(V|W,Y_2)\big) +\lambda_1(h(U)-h(U|Y_1,V,W)) \\
		=& \lambda_0\int  q(w)\big( d_W[Q](w) -\ln q(w)\big) {\rm d}w+\lambda_1\int  q(u) \big( 	d_U[Q](u)  \\&-\ln q(u)\big) {\rm d}u +\lambda_2 h(V),
	\end{align*} 
	where 
	\begin{align*}
		d_W[Q](w) 
		&=  \int q(u)q(v) p(y_1,y_2|v+u+w) \big( {\alpha}\ln Q(w|y_1) \\&\quad +{\bar{\alpha}}\ln Q(w|y_2) \big){\rm d}u{\rm d}v{\rm d}y_1{\rm d}y_2 ,\\ 
		d_U[Q](u)
		&= \int q(w)q(v) p(y_1,y_2|v+u+w) \big(\ln Q(u|v,\\&\quad\quad w,y_1)+\frac{\lambda_2}{\lambda_1}\ln Q(v|y_2,w)\big){\rm d}v{\rm d}w{\rm d}y_1{\rm d}y_2.
	\end{align*}
	Since $K_V$ is fixed, we omit the term $\lambda_2 h(V)$ in $F_C(q,Q)$ below.
	
	Similar to Theorem \ref{th1}, given the joint pdf $q(\cdot,\cdot,\cdot)$, the maximizing pdf $Q[q]$ of $F_C(q,\cdot)$ satisfies
	\begin{align*}	&Q[q](w|y_1)=g\big(w;K_W(K+\Sigma_1)^{-1}y_1,K_W-K_W(K\\& +\Sigma_1)^{-1} K_W\big):=g\big(w;\dot{A} y_1,\dot{W}_1\big),\\ &Q[q](w|y_2)=g\big(w;K_W(K+\Sigma_2)^{-1}y_2,K_W-K_W(K\\&  +\Sigma_2)^{-1}K_W\big):=g\big(w;\dot{B}y_2,\dot{W}_2\big),\\ &Q[q](v|y_2,w)=g\big(v;K_V(K_U+K_V+\Sigma_2)^{-1}(y_2-w),K_V\\&-K_V(K_U+K_V+\Sigma_2)^{-1}K_V\big):=g\big(v;\dot{C}(y_2-w),\dot{W}_3\big),\\  &Q[q](u|y_1,v,w)=g\big(u;K_U(K_U+\Sigma_1)^{-1}(y_1-v-w),K_U\\&-K_U(K_U+\Sigma_1)^{-1}K_U\big):=g\big(u;\dot{D}(y_1-v-w),\dot{W}_4\big).
	\end{align*}		
	Similar to Theorem \ref{th3}, under the assumption that $K_U,K_V,K_W \succ 0$, we get
	\begin{align*} 
		F_C\left(q,Q[q]\right) = & \lambda_0\int  q(w)\left( - \frac12 w^T D_W w -\ln q(w)\right) {\rm d}w\\&+\lambda_1\int  q(u) \left( 	- \frac12 u^T D_U u  -\ln q(u)\right) {\rm d}u,
	\end{align*}
	where $D_W = \alpha(\dot{W}^{-1}_1-\dot{A}^T\dot{W}^{-1}_1-\dot{W}^{-1}_1\dot{A})+\bar{\alpha}(\dot{W}^{-1}_2-\dot{B}^T\dot{W}^{-1}_2-\dot{W}^{-1}_2\dot{B})$ and $D_U = \dot{W}^{-1}_4-\dot{D}^T\dot{W}^{-1}_4-\dot{W}^{-1}_4\dot{D}+\dot{D}^T\dot{W}^{-1}_4\dot{D}+{\lambda_2}/{\lambda_1}(\dot{C}^T\dot{W}^{-1}_3\dot{C})$.
	Now, fixing $Q[q]$, we consider the following optimization problem:
	\begin{equation}
		\begin{aligned}
			\max_{\tilde{q}(\cdot,\cdot)}& \ \lambda_0\int  \tilde{q}(w)\left( - \frac12 w^T D_W w -\ln \tilde{q}(w)\right) {\rm d}w+\lambda_1\int  \tilde{q}(u) \notag\\& \ \left( 	- \frac12 u^T D_U u  -\ln \tilde{q}(u)\right) {\rm d}u\\
			{\rm s.t.}&
			\ \mathbb E_{\tilde{q}(u,w)}[UU^T+WW^T] = K-K_V.  
		\end{aligned}
	\end{equation} 
	Similar to Theorem \ref{th2}, let $\Gamma$ be chosen such that  $D_W + \Gamma/\lambda_0 \succ 0$, $D_U + \Gamma/\lambda_1 \succ 0$, and $(D_U + \Gamma/\lambda_1)^{-1} + (D_W + \Gamma/\lambda_0)^{-1} = K-K_V$. Then, the maximizing pdf $\tilde{q}[Q[q]]$ of $F_C(\cdot,Q[q])$ satisfies 
	\begin{align}
		\tilde{q}\big[Q[q]\big](w)&=g\left(w; 0, (D_W + \Gamma/\lambda_0\right)^{-1}),\\
		\tilde{q}\big[Q[q]\big](u)&=g\left(u; 0, (D_U + \Gamma/\lambda_1\right)^{-1}).
	\end{align}	
	Following the derivation of the GAB-P algorithm, we further obtain
	$$K_W^{-1} = D_W +\Gamma/\lambda_0,\quad  K_U^{-1} =  D_U + \Gamma /\lambda_1,$$ 
	where 
	\begin{align*} D_U=&(K_U\Sigma_1^{-1}K_U+K_U)^{-1}+\frac{\lambda_2}{\lambda_1}(K_U+\Sigma_2)^{-1}K_V\\&(K_U+K_V+\Sigma_2)^{-1}, 
	\end{align*} 
	$$D_W =K_W^{-1}-\alpha(K_U+K_V+\Sigma_1)^{-1}-{\bar{\alpha}}(K_U+K_V+\Sigma_2)^{-1},$$
	and  
	\begin{align*}
		\Gamma =& \lambda_0(K_W^{-1}-D_W) = \lambda_0(\alpha(K_U+K_V+\Sigma_1)^{-1}\\&+{\bar{\alpha}} (K_U+K_V+\Sigma_2)^{-1}).
	\end{align*} 
	Based on the analysis above and adapting the GAB-P algorithm in Section \ref{faal}, we obtain Algorithm \ref{Alg:BA214} for solving the optimization problem (\ref{vop}).
	
	\bibliography{IEEEabrv,BArefMY}

\end{document}